\documentclass[
  aps,prl,reprint,superscriptaddress,
  floatfix,
  amsmath,amssymb  
]{revtex4-2}

\usepackage{amsmath,amssymb}
\usepackage{braket}
\usepackage{graphicx}
\let\origincludegraphics\includegraphics
\renewcommand{\includegraphics}[2][]{%
  \IfFileExists{#2}{\origincludegraphics[#1]{#2}}{%
    \fbox{\parbox[c][0.18\textheight][c]{0.92\linewidth}{\centering Missing figure file\\\texttt{\detokenize{#2}}}}%
  }%
}
\usepackage{CJKutf8}
\usepackage{tikz}
\usepackage{placeins}
\usepackage{xcolor}
\usepackage{enumerate}
\usepackage{tikz}
\usepackage{xcolor}
\usepackage[
  colorlinks=true,
  linkcolor=blue,
  citecolor=blue,
  urlcolor=blue,
  breaklinks=true,
  pdftitle={Probing the Error-Mitigation Threshold with Matrix Product States},
  pdfauthor={J. Zhao et al.}
]{hyperref}

\usepackage{braket}
\usepackage{graphicx}

\usepackage{bm}  
\usepackage{tikz}
\usepackage{xcolor}
\usepackage{microtype}
\definecolor{spinA}{HTML}{0072B2} 
\definecolor{spinB}{HTML}{D55E00} 
\definecolor{blockA}{HTML}{0072B2} 
\definecolor{blockB}{HTML}{D55E00}
\definecolor{blockU}{HTML}{0072B2}     
\definecolor{blockUstar}{HTML}{D55E00} %

\usepackage{cleveref}
\crefname{equation}{Eq.}{Eqs.}
\crefname{figure}{Fig.}{Figs.}

\newcommand{\soptitle}{Probing the Error-Mitigation Threshold with Matrix Product States}

\crefname{table}{Table}{Tables}  

\begin{document}
\raggedbottom
\title{\soptitle}
\author{Jia-Yao Zhao (\begin{CJK*}{UTF8}{gbsn}赵家瑶\end{CJK*})}
\affiliation{Joint Quantum Institute and Joint Center for Quantum Information and Computer Science,\\
University of Maryland, College Park, Maryland 20742, USA}
\author{Zhi-Yuan Wei (\begin{CJK*}{UTF8}{gbsn}魏志远\end{CJK*})}
\email{zywei@umd.edu}
\affiliation{Joint Quantum Institute and Joint Center for Quantum Information and Computer Science,\\
University of Maryland, College Park, Maryland 20742, USA}
\author{Michael J.\ Gullans}
\email{mgullans@umd.edu}
\affiliation{Joint Quantum Institute and Joint Center for Quantum Information and Computer Science,\\
University of Maryland, College Park, Maryland 20742, USA}
\affiliation{National Institute of Standards and Technology, Gaithersburg, MD 20899, USA.}

\begin{abstract}
Quantum error mitigation relies on accurate noise characterization, but mismatches between the actual and characterized noise can be amplified and drive a sharp threshold between successful and failed mitigation. In random circuits, this threshold maps onto a random-field Ising transition, but previous exact numerics were limited to small one-dimensional and all-to-all systems, leaving explicit two-dimensional architectures unresolved. We develop a fixed-bond-dimension matrix-product-state method for the replicated transfer dynamics that extends threshold calculations beyond exact propagation while retaining the finite-size signatures of the transition. At system sizes beyond previous exact studies, we recover the predicted absence of a threshold for quenched disorder in 1D, obtain a sharper annealed all-to-all critical point, and resolve architecture-dependent finite-depth thresholds in 2D square and heavy-hex circuits. These results establish replicated tensor-network dynamics as a practical tool for probing error-mitigation thresholds in large and higher-dimensional noisy circuits.
\end{abstract}

\maketitle

\textit{Introduction}---
Quantum error mitigation~\cite{temme2017error, li2017efficient, cai2023quantum, endo2018practical} uses classical post-processing to extend the reach of noisy intermediate-scale quantum (NISQ) devices~\cite{preskill2018quantum,kim2023evidence}. Several powerful protocols, including probabilistic error cancellation (PEC)~\cite{temme2017error, van2023probabilistic, gupta2024probabilistic} and tensor-network error mitigation (TEM)~\cite{filippov2023scalable}, require an accurate
characterization of the noise. Small discrepancies between the true and
characterized noise can be amplified by the mitigation procedure as circuit size
and depth increase, raising a basic question for the scalability of near-term
quantum computation: when is error mitigation stable to imperfect noise
knowledge, and when does it break down?

\begin{figure}[!tb]
  \centering
  \includegraphics[width=\columnwidth]{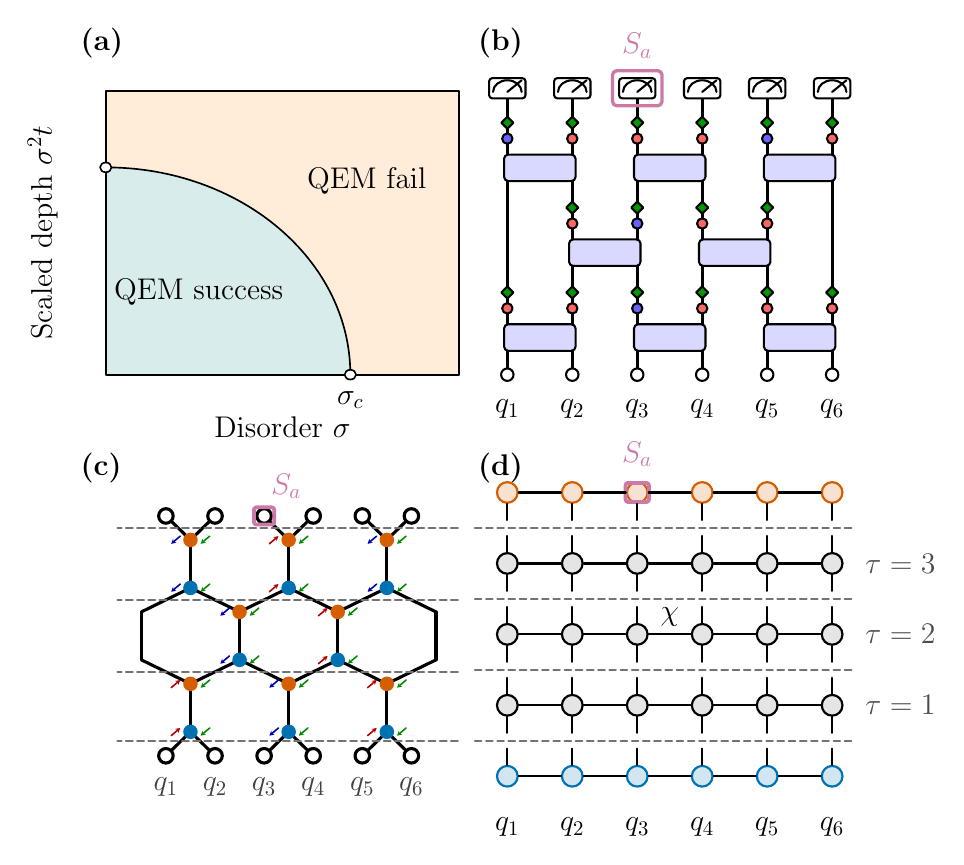} 
\caption{Statistical mapping and tensor-network evaluation of the error-mitigation threshold.
(a) Schematic phase diagram as a function of disorder strength and scaled depth $\sigma^2t$, separating QEM-success and QEM-failure regimes.
(b) 1D brickwork circuit with Haar-random two-qubit gates (purple boxes), local noise of two different rates (red/blue circles), and mean-field antinoise (green diamonds).
(c) $(1+1)$D two-replica Ising-like statistical model obtained from (b) after the local two-design average. Edges represent the resulting two-site interactions. Vertex colors distinguish the two replica-spin
sectors. Red/blue arrows denote the noise-induced fields for the two
rates, and green arrows denote the mean-field antinoise field, following the color convention in (b). Dashed lines separate transfer-matrix layers.
(d) Tensor-network contraction of the model in (c). Configuration weights are represented as an MPS. At each step, the transfer-matrix layer is applied to the MPS as an MPO, and the resulting MPS is truncated to bond dimension at most $\chi$. 
Across panels (b)--(d), pink outlines mark the final-time
evaluation of the central-qubit R\'enyi-2 entropy $S_a$ in each representation.}

  \label{fig:circuits}
\end{figure}

Recent theory~\cite{niroula2025error} predicts that this breakdown can take
the form of an error-mitigation threshold [cf.~Fig.~\ref{fig:circuits}(a)]. In noisy random circuits, imperfect
noise characterization acts as disorder in the mitigated dynamics, which admits
a statistical-mechanics description closely related to a random-field Ising
model (RFIM)~\cite{imry1975random,nattermann1998theory,niroula2025error}. The resulting disorder-driven transition separates a regime in which mitigation remains stable to depths growing with system size from one in which the mitigated signal fails on average. Whether this transition exists depends on the spatial dimension $D$
and the temporal correlations of the noise mismatch. Prior theory predicts no sharp transition for one-dimensional circuits with time-independent, or quenched disorder, but a nontrivial threshold for circuits with spacetime-fluctuating, or annealed, disorder in
$D\ge2$ and for circuits with quenched disorder in
$D\ge3$~\cite{imry1975random,nattermann1998theory,
niroula2025error}. The marginal cases, annealed disorder in $D=1$ and quenched disorder in $D=2$, are not considered here.

Previous exact simulations were limited to 1D and all-to-all circuits with $N\lesssim20$~\cite{niroula2025error}, leaving threshold predictions for  2D architectures as an open problem. This limitation is particularly consequential because square-grid and heavy-hex connectivity underpin leading superconducting processors and recent random-circuit and error-mitigation experiments~\cite{arute2019quantum,
acharya2023suppressing,kim2023evidence,hetenyi2024creating}. Direct matrix product state (MPS) evolution does not automatically remove the bottleneck: mapping a 2D lattice to a 1D ordering generates long-range updates, while full-state accuracy can require a bond dimension that grows rapidly with lattice width and evolution depth~\cite{orus2014practical,perez2006matrix,
schollwock2011density,mcculloch2007density}. The challenge is therefore to compress the replicated dynamics sufficiently to reach 2D scales while retaining the finite-size signatures that locate the transition.

In this Letter, we address this challenge with a fixed-bond-dimension MPS method for the two-replica transfer dynamics of error-mitigated random circuits~\cite{zhou2019emergent,nahum2017quantum, nahum2018operator,niroula2025error}. The method reaches the deep-circuit regime needed to resolve the transition. We first formulate the circuit-averaged second moments as a transfer-matrix evolution in the effective Ising representation, and then implement this evolution as an MPS/MPO contraction with constant bond-dimension truncation. As benchmarks, we revisit one-dimensional circuits with quenched disorder and all-to-all circuits with annealed disorder, recovering the expected absence of a threshold in the former case and obtaining a sharper critical point in the latter~\cite{niroula2025error}. We then apply the same protocol to square and heavy-hex lattices, where crossings of the local R\'enyi-2 entropy and mutual information yield quantitative, architecture-resolved finite-depth phase boundaries.

\textit{Setup}---We study noisy Haar-random circuits on $N$ qubits evolved for $t$ complete entangling layers. Each layer $\ell=1,\ldots,t$ is decomposed into $n_{\rm step}$ disjoint-gate substeps $\mu=1,\ldots,n_{\rm step}$, with $n_{\rm step}=2,4,1,5$ for the 1D brickwork, 2D square, all-to-all, and 2D heavy-hex geometries, respectively. We label an elementary transfer step by $\tau=(\ell,\mu)$. After the disjoint Haar gates at step $\tau$, we apply a product noise channel $\mathcal{E}_\tau=\bigotimes_x\mathcal{E}_{x,\tau}$ followed by a product mean-field antinoise map $\mathcal{A}_\tau=\bigotimes_x\mathcal{A}_{x,\tau}$ that represents the application of quantum error mitigation.  The local forms of these maps are specified below. The $L\times L$ square and $L_x\times L_y$ heavy-hex~\cite{hetenyi2024creating} lattices use cylindrical boundary conditions, periodic in one direction and open in the other; gate schedules and boundary conditions are given in the Supplemental Material (SM)~\cite{supp}.

The single-qubit noise channel acting on qubit \(x\) at time step \(\tau\) is
\begin{equation}
\mathcal{E}_{x,\tau}(\rho)=(1-q_{x,\tau})\,\rho+q_{x,\tau}\,\mathrm{Tr}(\rho)\frac{\mathbb{I}}{2},
\end{equation}
where the depolarizing rate $q_{x,\tau}\in\{q_1,q_2\}$, with $q_1<q_2$, is drawn from a bimodal distribution parametrized by $f$, with $\Pr(q_{x,\tau}=q_1)=f$ and $\Pr(q_{x,\tau}=q_2)=1-f$. This minimal model captures spacetime fluctuations in the mismatch
between the actual local noise and the uniform noise model used for mitigation. We fix the mean error rate $\bar q=fq_1+(1-f)q_2=0.1$ and characterize the disorder strength as
\begin{equation}
\sigma=\sqrt{f(1-f)}\,|q_2-q_1|.
\label{eq:disorder}
\end{equation}

We use annealed disorder, with rates $q_{x,\tau}$ drawn independently
at each spacetime location $(x,\tau)$, and quenched disorder, with the pattern
over one complete layer drawn once and repeated across layers. These ensembles represent limiting cases of temporal correlations in the noise-characterization mismatch.

To model mitigation of the characterized noise, we introduce the uniform mean-field antinoise map~\cite{niroula2025error}
\begin{equation}
\mathcal{A}_{x,\tau}(\rho) = \eta\,\rho + (1-\eta)\mathrm{Tr}(\rho)\,\frac{\mathbb{I}}{2} ,
\qquad
\eta = \frac{1}{1-q_a}.
\end{equation}
This formal inverse of a depolarizing channel with rate $q_a$ represents the inverse-noise operation implemented through error-mitigation postprocessing rather than a physical channel. We choose $q_a$ so that noise and antinoise cancel in the mean logarithmic decay rate, giving the zero-mean-field condition,
\begin{equation}
1-q_a = (1-q_1)^f (1-q_2)^{1-f}.
\label{eq:MF}
\end{equation}
Combining these ingredients, the full evolution for one step is  \(\rho \mapsto \mathcal{A}_\tau\circ \mathcal{E}_\tau \circ \mathcal{U}_\tau(\rho)\) [Fig.~\ref{fig:circuits}(b)], where \(\mathcal{U}_\tau = \bigotimes_{\langle x,y \rangle \in E_\tau} \mathcal{U}_{x,y}\) applies independent two-qubit Haar-random unitary gates to the disjoint pairings \(E_\tau\) dictated by the circuit geometry at step $\tau$ ~\cite{supp}.

\textit{Local probe}---Our diagnostics for the error-mitigation threshold are circuit-averaged R\'enyi-2 observables~\cite{zhou2019emergent}. Throughout, logarithms are taken to base two. For a subsystem $a$, let $\rho_a$ denote its reduced density matrix. Using the swap trick,
$\mathrm{Tr}\!(\rho_a^2)\!=\!\mathrm{Tr}\!(\rho_a^{\otimes2}\!\bigotimes_{x\in a}\!\mathrm{SWAP}_x)$, where $\mathrm{SWAP}_x$ swaps the two replicas on qubit $x$, we evaluate the averaged R\'enyi-2 entropy as
\begin{equation}
S_a=-\log\!\left[\mathrm{Tr}\!\left(\mathbb{E}_{U}\!\left[\rho_a^{\otimes 2}\right]\bigotimes_{x\in a}\mathrm{SWAP}_x\right)\right],
\label{eq:Sa}
\end{equation}
and for subsystems $a,b$ the corresponding mutual information $I_{ab}=S_a+S_b-S_{ab}$~\cite{niroula2025error,supp}. The replicas share the same noisy, error-mitigated circuit realization.

\textit{Statistical mapping}---We summarize the two-replica transfer-matrix mapping; see SM for more details~\cite{supp}. After Haar averaging, the replicated density matrix $\bar\rho^{(2)}_\tau \equiv \mathbb{E}_U[\rho_\tau^{\otimes 2}]$, where $\rho_\tau$ is the density matrix at step $\tau$,  closes on the operator basis $\{\mathbb{I}_4,S\}^{\otimes N}$. $\mathbb{I}_4$ is the identity on the two-replica single-qubit space and $S$ swaps the two replicas. We therefore expand
\begin{equation}
\bar\rho^{(2)}_\tau
= \sum_{\alpha \in \{0,1\}^N} P_\tau(\alpha)\;
\bigotimes_{x=1}^{N} \mathbb{I}_4^{\,1-\alpha_x} S^{\,\alpha_x},
\label{eq:rho2_expansion_prl}
\end{equation}
where the coefficients $P_\tau(\alpha)$ define configuration weights for Ising-like variables $\alpha_x\in\{0,1\}$ on the circuit spacetime lattice~\cite{ware2023sharp,nelson2025error}; see Fig.~\ref{fig:circuits}(c)
for a representative mapping. Within this averaged-observable two-design description, PEC and TEM generate the same dressed transfer process for the R\'enyi-2 observables considered here; see the End Matter.

Treating the coefficients $P_\tau(\alpha)$ from Eq.~\eqref{eq:rho2_expansion_prl} as components of a coefficient vector $|P_\tau\rangle$, the system evolves via a linear transfer process
\begin{equation}
|P_{\tau+1}\rangle
=
\mathbb{M}_\tau\,|P_\tau\rangle,
\qquad
\mathbb{M}_\tau \equiv M[\mathcal{A}_\tau]\;M[\mathcal{E}_\tau]\;M[\mathcal{U}_\tau].
\label{eq:transfer_update}
\end{equation}
Here \(\mathbb{M}_\tau\) is the full transfer matrix for time step \(\tau\) acting on the operator basis \(\{\mathbb{I}_4,S\}^{\otimes N}\), obtained by composing the single-site noise and antinoise product maps, \(M[\mathcal{E}_\tau]=\bigotimes_x M[\mathcal{E}_{x,\tau}]\) and \(M[\mathcal{A}_\tau]=\bigotimes_x M[\mathcal{A}_{x,\tau}]\), with the Haar-averaged gate update \(M[\mathcal{U}_\tau] = \bigotimes_{\langle x,y \rangle \in E_\tau} M_{\mathrm{Haar}}\) applied to the disjoint pairings \(E_\tau\) active at that step. Specifically, a Haar-averaged two-qubit gate induces a local update on the ordered two-site basis $\{\mathbb{I}_4\mathbb{I}_4,\mathbb{I}_4S,S\mathbb{I}_4,SS\}$ given by
\begin{equation}
M_{\mathrm{Haar}}
=
\begin{pmatrix}
1 & \tfrac{2}{5} & \tfrac{2}{5} & 0 \\
0 & 0 & 0 & 0 \\
0 & 0 & 0 & 0 \\
0 & \tfrac{2}{5} & \tfrac{2}{5} & 1
\end{pmatrix}.
\label{eq:Mhaars}
\end{equation}

Single-qubit channels act independently on each site and preserve
the span $\{\mathbb{I}_4,S\}$. The corresponding local update matrices
$M[\mathcal{E}_{x,\tau}]$ and $M[\mathcal{A}_{x,\tau}]$ are derived in
Ref.~\cite{niroula2025error} and the SM~\cite{supp}.

Physically, the transfer process maps the $D$-dimensional circuit to a $(D{+}1)$-dimensional RFIM on spacetime sites $(x,\tau)$. Haar-averaged two-qubit gates induce local couplings set by the gate
pattern, while the sampled rates $q_{x,\tau}$ generate a spacetime-dependent random field. The antinoise $\mathcal{A}_{x,\tau}$ contributes a uniform on-site bias fixed by Eq.~\eqref{eq:MF}, thereby canceling the mean noise-induced field.

\begin{figure}[t]
  \centering
  \includegraphics[width=\linewidth]{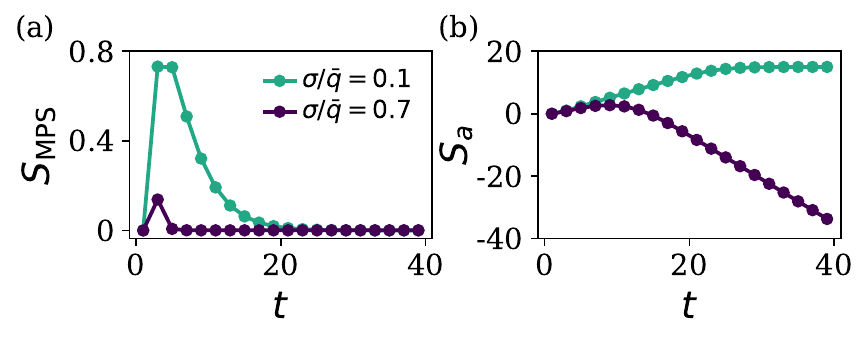}
  \caption{
Entanglement evolution in the annealed-disorder all-to-all statistical model for $N=30$.
(a) Bipartite von Neumann entanglement entropy $S_{\mathrm{MPS}}(t)$ of the normalized MPS across a fixed cut, which diagnoses the entanglement cost of the MPS contraction.
(b) Corresponding circuit-averaged bipartite R\'enyi-2 entropy $S_a(t)$
[Eq.~\eqref{eq:Sa}] for the subsystem on one side of that cut, used to probe the error-mitigation threshold.
}
  \label{fig:1d_S_MPS_vertical}
\end{figure}

\textit{Fixed-bond-dimension MPS}---
Estimating the error-mitigation threshold requires contracting the \((D+1)\)-dimensional spacetime tensor network over many entangling layers and disorder realizations. Exact evolution is exponentially costly, while 2D gate schedules induce long-range MPS updates and rapidly growing bond dimensions. We therefore seek to preserve the mitigation threshold signatures without requiring global convergence of the full coefficient vector. 

We represent the coefficient vector $|P_\tau\rangle$ associated with Eq.~\eqref{eq:rho2_expansion_prl} as a MPS over binary variables $\alpha_x\in\{0,1\}$~\cite{perez2006matrix}:
\begin{equation}
|P_\tau\rangle
=
\sum_{\alpha\in\{0,1\}^{N}}
A^{\alpha_1}_1(\tau)\cdots A^{\alpha_N}_N(\tau)\,|\alpha_1\cdots\alpha_N\rangle.
\label{eq:Ptau_MPS}
\end{equation}

Here $|P_\tau\rangle$ encodes configuration weights of the effective statistical model. The transfer matrix $\mathbb{M}_\tau$ in Eq.~\eqref{eq:transfer_update} acts linearly on $|P_\tau\rangle$ and is implemented as a matrix product operator~\cite{hubig2017generic}, as illustrated in
Fig.~\ref{fig:circuits}(d). We apply $\mathbb{M}_\tau$ using Time-Evolving Block Decimation (TEBD)~\cite{orus2008infinite,paeckel2019time,urbanek2016parallel} and truncate the resulting MPS to maximum bond dimension $\chi$
after each step $\tau$~\cite{vidal2004efficient}.  Fixing $\chi$ prevents the bond dimension from growing with circuit depth and defines a compressed transfer evolution on the finite-$\chi$ MPS manifold. The exact finite-system evolution is recovered at sufficiently large $\chi$. At accessible bond dimensions, we therefore locate the transition independently at each $\chi$ and use the stability of the extracted critical parameters as a finite-$\chi$ saturation test.

The efficiency of this representation is controlled by entanglement
in the coefficient MPS $|P_{\tau}\rangle$. We diagnose this entanglement using the $\ell_2$-normalized coefficient vector $|\tilde{P}_\tau\rangle\equiv|P_\tau\rangle/\|P_\tau\|_2$. For a subsystem $R$ in the chosen MPS ordering, $S_{\mathrm{MPS}}(\tau)$ is the von Neumann entropy of its reduced density matrix, defined explicitly in the SM~\cite{supp}. As a representative illustration, $S_{\mathrm{MPS}}$ remains small throughout the evolution in the annealed-disorder all-to-all geometry [Fig.~\ref{fig:1d_S_MPS_vertical}(a)].

The observed compressibility can be understood from the asymptotic structure of the transfer dynamics. In the ideal circuit, the Haar-averaged two-site update
[Eq.~\eqref{eq:Mhaars}] leaves the local $\mathbb I_4\mathbb I_4$ and $SS$
sectors invariant while projecting mixed local sectors into them. Thus the long-time ideal coefficient vector lies in the span of the
uniform identity and uniform swap product configurations, $\mathbb I_4^{\otimes N}$ and $S^{\otimes N}$, and therefore admits
an exact MPS representation with bond dimension at most two. In the mitigated noisy dynamics, noise and antinoise redistribute weight among configurations with different Hamming weight $|\alpha|$. Below threshold, residual noise favors identity-like configurations with small $|\alpha|$, whereas above threshold over-mitigation can amplify swap-rich configurations with large $|\alpha|$. 

In the same calculation, the circuit-averaged R\'enyi-2 entropy $S_a$ distinguishes the two regimes: for weak disorder it saturates at late times, whereas for strong disorder it decreases and can become negative due to over-compensation by the antinoise map, signaling the breakdown of error mitigation [Fig.~\ref{fig:1d_S_MPS_vertical}(b)]. Thus the coefficient vector can remain computationally compressible while the extracted second-moment observables retain the disorder dependence needed to diagnose the transition.

Although exact contractions are basis invariant, finite-$\chi$ SVD truncation is path dependent: local basis rescalings change the singular-value spectrum and can alter which Schmidt sectors are retained. We therefore evolve $|P_\tau\rangle$ in the unrescaled $\{\mathbb I_4,S\}^{\otimes N}$ basis without post-truncation normalization. This prescription retains disorder sensitivity at substantially smaller $\chi$ than the locally normalized-basis protocol with post-truncation renormalization. Comparisons of the two protocols and analysis of the associated weak-disorder artifact are given in the End Matter and SM~\cite{supp}.

\begin{figure}[t]
  \centering
  \includegraphics[width=\columnwidth]{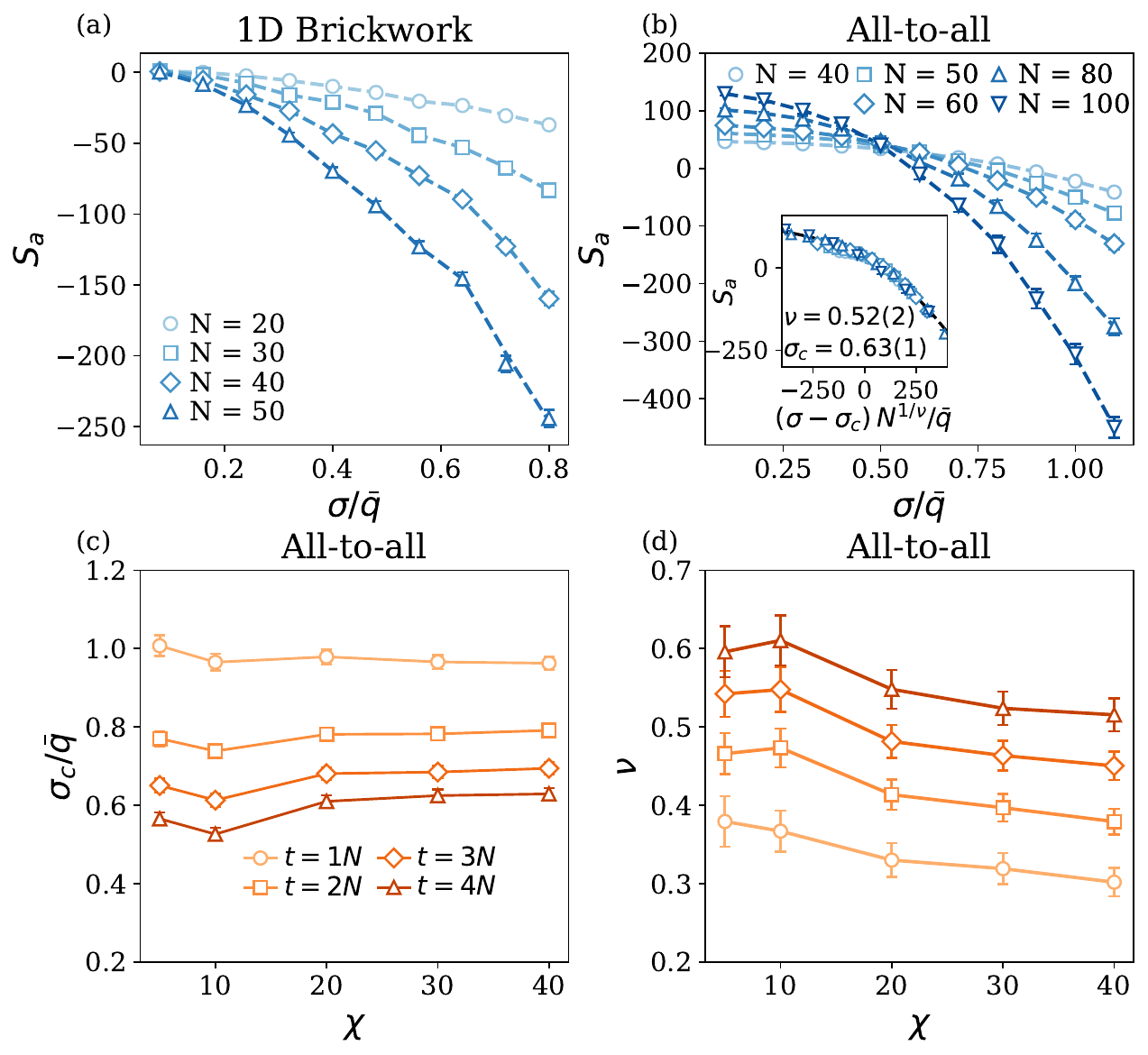}
\caption{Benchmark of the fixed-bond-dimension MPS method.
(a),(b) Circuit-averaged single-site R\'enyi-2 entropy $S_a$ at
$t=4N$ and $\chi=40$ for (a) quenched 1D and (b) annealed
all-to-all circuits. Colored lines are guides to the eye; the inset
in (b) shows the finite-size-scaling collapse, with the dashed black
curve denoting the fitted scaling function.
(c),(d) Extracted $\sigma_c/\bar q$ and $\nu$ versus $\chi$ for the
all-to-all geometry. The parameter and inset uncertainties are
covariance estimates from the nonlinear collapse fits.}
  \label{fig:crit_vs_chi}
\end{figure}

\begin{figure}
  \centering
  \includegraphics[width=\linewidth]{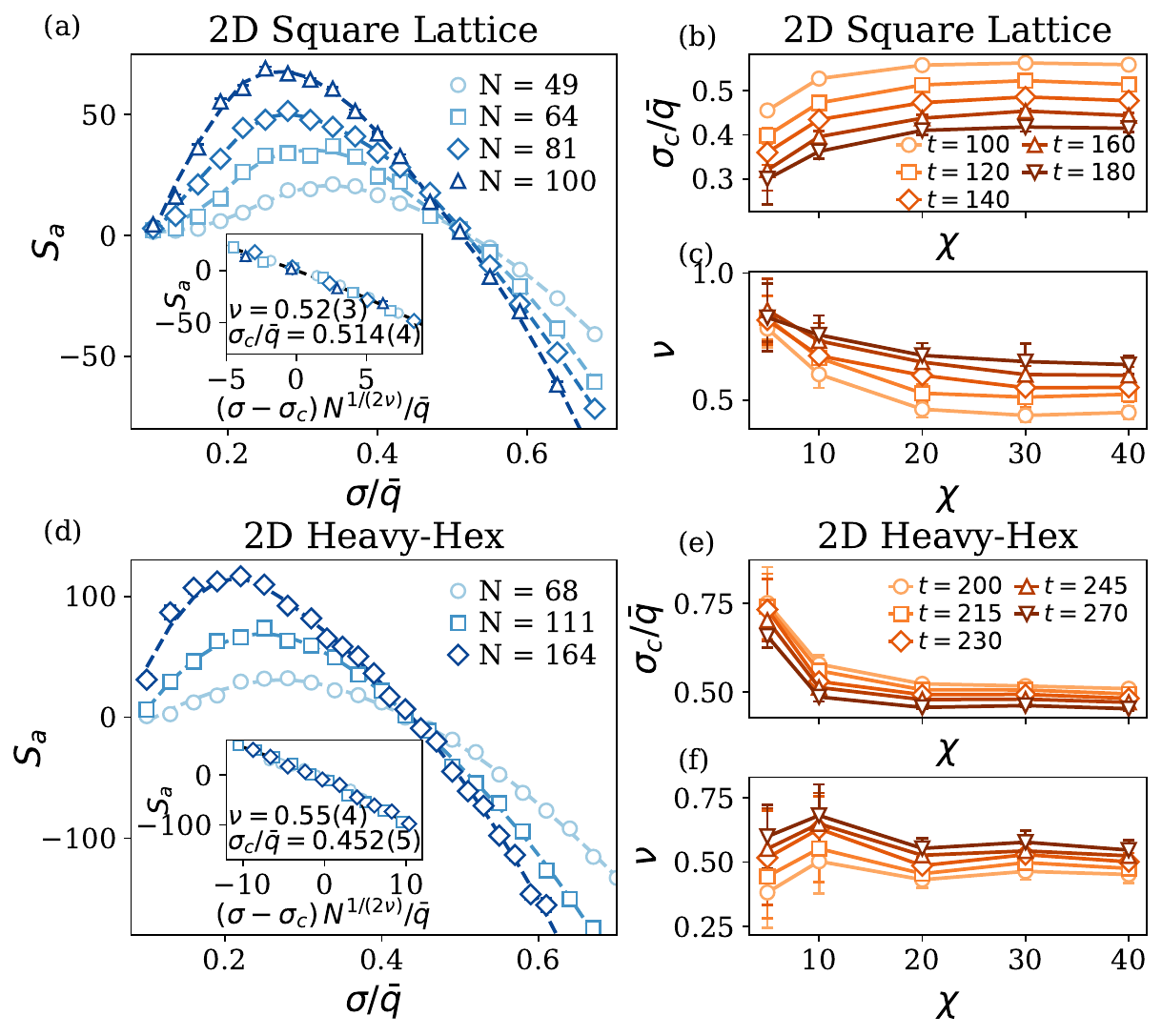}
\caption{Finite-depth thresholds in 2D architectures.
(a),(d) Circuit-averaged $S_a$ for annealed disorder at $\chi=40$:
(a) the square lattice at $t=120$ and (d) the heavy-hex lattice at $t=270$; insets show finite-size collapses with dashed black fitted scaling functions.
(b),(c) Extracted $\sigma_c/\bar q$ and $\nu$ versus $\chi$ for the square lattice, and (e),(f) for heavy-hex. The parameter and inset uncertainties are covariance estimates from the nonlinear collapse
fits.}
 \label{fig:threshold_Sa_Iab}
\end{figure}

\textit{Numerical benchmarks}---Prior work predicts no nonzero-disorder mitigation threshold for quenched 1D circuits but a finite threshold for annealed all-to-all circuits~\cite{niroula2025error}. We test these predictions with fixed-$\chi$ MPS by evaluating the circuit-averaged central-qubit R\'enyi-2 entropy $S_a$ [Eq.~\eqref{eq:Sa}] after $t=N,2N,3N,4N$ layers. We fix the mean depolarizing rate to $\bar q=0.1$ and the probability of drawing the lower rate $q_1$ to $f=0.9$, so scanning the disorder strength $\sigma/\bar q$ uniquely determines $q_1$ and $q_2$. For the quenched 1D geometry shown schematically in Fig.~\ref{fig:circuits}(b), the representative $\chi=40$ data in Fig.~\ref{fig:crit_vs_chi}(a) show no crossing for $\sigma/\bar q>0$ over the tested ranges of system size $N$ and layers $t$. This conclusion is unchanged for all tested bond dimensions, $\chi\in\{5,10,20,30,40\}$.

In contrast, annealed all-to-all circuits display a finite-size crossing in $S_a$ [Fig.~\ref{fig:crit_vs_chi}(b)], marking the
mitigation threshold. Its location near $S_a=0$ reflects the change from the residual-noise-dominated ordered phase (successful mitigation, $S_a>0$) to the over-mitigated disordered phase ($S_a<0$). The inset shows a good finite-size collapse at $\chi=40$. Repeating the analysis across bond dimensions gives the threshold location and exponent in Figs.~\ref{fig:crit_vs_chi}(c),(d). $\sigma_c/\bar q$ shows only weak dependence on the bond dimension $\chi$, whereas $\nu$ is more sensitive to finite-$\chi$ truncation but changes little by $\chi=40$. We therefore quote the largest-bond-dimension result at $\chi=40$ as $\sigma_c/\bar q=0.63(1)$, consistent with the prior mean-field and small-system entropy estimate $0.65(5)$~\cite{niroula2025error}. Independent finite-size scaling of the mutual information $I_{ab}$ yields $\sigma_c/\bar q=0.63(1)$, consistent with the previously reported $I_{ab}$ range $0.55(5)$--$0.65(5)$~\cite{niroula2025error}; details are given in the SM~\cite{supp}. These results validate fixed-$\chi$ MPS for studying error-mitigation thresholds at system sizes beyond the reach of exact propagation~\cite{niroula2025error}.

\textit{2D thresholds}---
We next apply the protocol to the annealed-disorder 2D architectures implemented with the fixed gate schedules~\cite{supp}. These geometries are motivated by superconducting platforms, including square-grid layouts used in Google's random-circuit experiments and heavy-hex connectivity used in IBM processors~\cite{arute2019quantum,kim2023evidence,hetenyi2024creating}.  Their two-replica mappings produce computationally challenging $(2+1)$-dimensional spacetime tensor networks, which we contract across the full 2D lattice and many time layers. Although RFIM arguments predict a nontrivial mitigation threshold for annealed disorder in two dimensions~\cite{imry1975random, nattermann1998theory,niroula2025error}, this prediction has not been tested numerically for explicit 2D gate schedules. We therefore determine architecture-resolved finite-depth phase boundaries. Because the critical disorder depends on depth [Fig.~\ref{fig:circuits}(a)], each $t$ defines a distinct point $\sigma_c(t)$ on the corresponding
phase boundary.

For both architectures, $S_a$ exhibits nonzero-disorder crossings and finite-size collapses. Representative $\chi=40$ results are shown for the square lattice at $t=120$ [Fig.~\ref{fig:threshold_Sa_Iab}(a)] and the heavy-hex lattice at $t=270$ [Fig.~\ref{fig:threshold_Sa_Iab}(d)]. Across the depths shown, the extracted $\sigma_c/\bar q$ and $\nu$ are essentially saturated at the largest bond dimensions
[Figs.~\ref{fig:threshold_Sa_Iab}(b),(c),(e),(f)]. The representative finite-depth estimates are $\sigma_c/\bar q=0.514(4)$ and $\nu=0.52(3)$ for square, and $\sigma_c/\bar q=0.452(5)$ and $\nu=0.55(4)$ for heavy-hex.

In the weak-disorder regime, a secondary crossing near $\sigma/\bar q\simeq 0$ is visible at finite $\chi$ [cf.~Figs.~\ref{fig:threshold_Sa_Iab}(a,d)]. As summarized in the End Matter, we identify this feature as a finite-$\chi$ truncation artifact and extract the quoted thresholds from the stable larger-disorder crossing. At $\chi=40$, independent finite-size scaling analyses of the mutual information $I_{ab}$ yield threshold locations consistent with the corresponding $S_a$-based estimates for the square and heavy-hex lattices [Figs.~\ref{fig:threshold_Sa_Iab}(b) and
\ref{fig:threshold_Sa_Iab}(e), respectively]; details are given in the SM~\cite{supp}.

\textit{Conclusion}---We developed a fixed-bond-dimension tensor-network approach for the two-replica transfer dynamics of PEC, with the same local two-design description applying to TEM. After reproducing the expected threshold behavior in the previously studied quenched 1D and annealed all-to-all limits~\cite{niroula2025error}, we used the method to resolve quantitative finite-depth thresholds and critical scaling for computationally demanding 2D square and heavy-hex architectures at system sizes beyond exact propagation. Extending the framework to more realistic noise models and circuit ensembles could enable direct experimental observation of error-mitigation threshold behavior.

\begin{acknowledgements}
\emph{Acknowledgements}.--We thank A. Seif and Z. Minev for helpful discussions.  Research supported in part by NSF QLCI award OMA2120757. This work was performed in part at the Kavli Institute for Theoretical Physics (KITP), which is supported by grant NSF PHY-2309135. Z.Y.W.~was supported in part by the NSF STAQ program (awards No.~1818914 and No.~2325080).
\end{acknowledgements}

\bibliography{reference_revised}

\appendix

\section{Two-design transfer-rule equivalence of TEM and PEC}
\label{app:tem_pec}
Within the two-design statistical mapping, tensor-network error mitigation (TEM)~\cite{filippov2023scalable} and probabilistic error cancellation (PEC)~\cite{temme2017error,endo2018practical} generate the same dressed local transfer rules. The significance of this observation is not that the two protocols are operationally identical, but that the replicated statistical model used here applies to both once the local noise/antinoise dressing is projected into the two-design commutant.

A convenient derivation uses a noisy--ideal overlap representation, in which one doubled copy follows the noisy evolution while the second follows the corresponding ideal evolution. This is an auxiliary representation for establishing the TEM/PEC equivalence at the local transfer-rule level. In this noisy--ideal representation, the PEC single-site updates are
\begin{align}
M_{\mathrm{NI}}\!\left[\mathcal{E}^{(1)}_{x,\tau}(q_{x,\tau})\right]
&=
\begin{pmatrix}
1 & \tfrac{q_{x,\tau}}{2}\\
0 & 1-q_{x,\tau}
\end{pmatrix},
\label{eq:MNI_noise_endmatter}
\\
M_{\mathrm{NI}}\!\left[\mathcal{A}^{(1)}_{x,\tau}(q_a)\right]
&=
\begin{pmatrix}
1 & \tfrac{1-(1-q_a)^{-1}}{2}\\
0 & (1-q_a)^{-1}
\end{pmatrix}.
\label{eq:MNI_antinoise_endmatter}
\end{align}
In TEM, the inverse-circuit correction can be commuted through the Hilbert--Schmidt contraction, so the ideal copy effectively experiences the transpose map of the local inverse-noise dressing. The relevant Haar average is then the mixed twirl
\begin{equation}
\mathcal{M}^{(2,*)}_{2}[O]
\equiv
\int dU\,(U\otimes U^{*})\,O\,(U^{\dagger}\otimes U^{T})
=
\Bigl(\mathcal{M}^{(2)}_{2}[O^{T_2}]\Bigr)^{T_2},
\label{eq:twirl_UUstar_endmatter}
\end{equation}
which closes on the same two-dimensional commutant sector as the usual two-copy twirl, with $S$ replaced by $\widetilde S\equiv S^{T_2}$. Here $T_2$ denotes partial transpose on the second copy. Consequently, the numerical Haar-gate transfer matrix is unchanged.

Acting on the ordered basis $\{\mathbb{I}_4,\widetilde S\}$, the TEM noisy--ideal single-site update matrix is
\begin{equation}
M_{\mathrm{NI}}^{\mathrm{(TEM)}}
=
\begin{pmatrix}
1 & \tfrac{1}{2}\!\left(1-\frac{1-q_{x,\tau}}{1-q_a}\right)\\
0 & \frac{1-q_{x,\tau}}{1-q_a}
\end{pmatrix}.
\label{eq:MNI_TEM_endmatter}
\end{equation}
Direct multiplication of the PEC noisy and antinoise matrices  in Eqs.~\eqref{eq:MNI_noise_endmatter} and~\eqref{eq:MNI_antinoise_endmatter}
yields exactly the same local update, so the dressed transfer process governing the threshold is identical for PEC and TEM within the two-design description. The threshold estimates extracted in the main text therefore apply to both protocols at this level of description.

\section{Finite-\texorpdfstring{$\chi$}{chi} truncation path and the weak-disorder artifact}
\label{app:trunc_path}

As discussed in the Letter, the finite-$\chi$ SVD truncation is not basis-invariant, making the truncation path a critical element of the numerical protocol. The choice of the unrescaled basis $\{\mathbb{I}_4,S\}^{\otimes N}$ is
motivated by the asymptotic two-replica state of an ideal Haar-random circuit:
\begin{equation}
\bar\rho^{(2)}_{\infty}
=
\frac{1}{2^N+4^N}\Bigl(\mathbb{I}_4^{\otimes N}+S^{\otimes N}\Bigr).
\label{eq:ideal_fp_nonorm_endmatter}
\end{equation}
In this unrescaled representation, the all-identity and all-swap configurations carry equal weight. By contrast, in the locally normalized basis
$\{\mathbb{I}_4/4,S/2\}^{\otimes N}$, the same fixed point is weighted toward the identity sector by a relative factor of $2^N$. At small $\chi$, this representation-dependent bias can suppress swap-rich sectors before physically relevant differences accumulate, delaying convergence. Evolving in the unrescaled basis and omitting post-truncation normalization preserves this exact balance at shallow depths and ensures that the overall norm retains its physical disorder dependence even after strong truncation.

The low-disorder crossing visible in the 2D data near $\sigma/\bar q\approx 0$ is a direct consequence of this finite-$\chi$  truncation. At zero disorder, $q_1=q_2=\bar q$, and the zero-mean-field condition
gives $q_a=\bar q$, so the noise and antinoise maps cancel exactly.
For a single-site subsystem $a$, Eq.~\eqref{eq:ideal_fp_nonorm_endmatter}
therefore gives
\begin{equation}
\begin{split}
S_a
&=
-\log
\operatorname{Tr}
\left[
\bar{\rho}^{(2)}_{\infty}\,
\mathrm{SWAP}_a
\right]
\\
&=
-\log
\left(
\frac{
2\,4^{N-1}+4\,2^{N-1}
}{
4^N+2^N
}
\right)
\\
&=
-\log
\left(
\frac{2^{N-1}+2}{2^N+1}
\right)
\\
&=
1-\log
\left(
1+\frac{3}{2^N+1}
\right)
=
1-\mathcal{O}(2^{-N}),
\end{split}
\label{eq:ideal_single_site_entropy_endmatter}
\end{equation}
where $\mathrm{SWAP}_a$ acts on site $a$ and as the identity on its
complement. Thus, near zero disorder, the exact finite-size separation
between the $S_a$ curves is exponentially small. A finite-$\chi$
truncation bias need not share this exponential suppression and can
therefore compete with the physical signal, reverse the finite-size
ordering, and generate an apparent crossing near
$\sigma/\bar q\simeq0$. This interpretation is also consistent with
prior exact small-system calculations of the all-to-all benchmark,
which exhibit the physical finite-disorder transition without an
analogous near-zero-disorder crossing~\cite{niroula2025error}. We
therefore interpret the weak-disorder feature as a finite-$\chi$
truncation artifact and extract the reported estimates from the
larger-disorder crossing, which shows substantially weaker
bond-dimension dependence.

\clearpage
\onecolumngrid
\section*{Supplemental Material}
\setcounter{equation}{0}
\renewcommand{\theequation}{S\arabic{equation}}
\setcounter{figure}{0}
\renewcommand{\thefigure}{S\arabic{figure}}
\setcounter{table}{0}
\renewcommand{\thetable}{S\arabic{table}}

\subsection*{Circuit geometries}
\label{sec:supp_circuit_geometries}
\begin{figure}[!htbp]
  \centering
  \includegraphics[width=\linewidth]{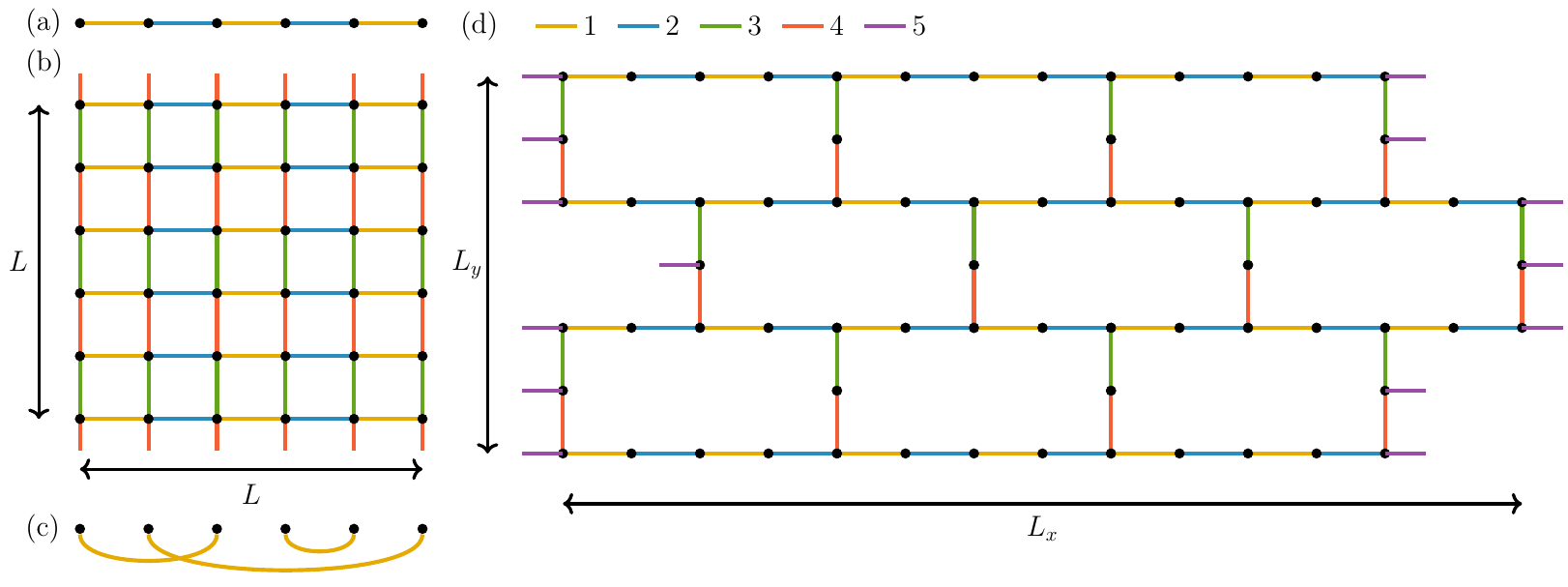}
  \caption{
  Gate schedules.
  (a) 1D brickwork geometry with two time steps per layer.
  (b) Square lattice with four time steps per layer.
  (c) All-to-all architecture with one random disjoint matching per layer.
  (d) heavy-hex patch lattice with five time steps per layer.
  Colors indicate disjoint two-qubit gate sets applied sequentially within one
  complete entangling layer. Boundary half-edges indicate wraparound couplings
  for cylindrical boundary conditions.
  }
  \label{fig:supp_gate_schedules}
\end{figure}

Figure~\ref{fig:supp_gate_schedules} shows the one-layer decompositions used
for the circuit geometries: 1D brickwork, square, all-to-all, and heavy-hex geometries. 
 A layer denotes one complete entangling cycle, decomposed into
architecture-dependent sets of disjoint two-qubit gates. For the square geometry, we use $L\times L$ patches with $N=L^2$. For the heavy-hex geometry, we use $3\times 3$, $4\times 4$, and $5\times 5$ patches; in the rectangular embedding these correspond to $(L_x,L_y)=(7,15)$, $(9,19)$, and $(11,23)$, with qubit numbers $N=68,111,164$, respectively.

\FloatBarrier
\FloatBarrier

\subsection*{Two-replica transfer process}
\label{app:mapping}

This section derives the local transfer rules of the two-replica statistical model~\cite{li2023entanglement,ware2023sharp,liu2024noise,gao2024limitations,zhou2019emergent} obtained from the second-moment Haar average.
Throughout, $\mathbb{I}_4$ denotes the identity on the two-replica single-qubit space $(\mathbb{C}^2)^{\otimes 2}$ and $S$ is the swap operator,
$
S\ket{\alpha}\ket{\beta}=\ket{\beta}\ket{\alpha},
$
which exchanges the two replicas. We work in the non-normalized operator basis $\{\mathbb{I}_4,S\}^{\otimes N}$.

Let $V\in U(d)$ and let $O$ be an operator on $(\mathbb{C}^d)^{\otimes 2}$. The second-moment Haar twirl is the linear map
\begin{equation}
\mathcal{M}^{(2)}_d[O]
\equiv
\int dV\, (V^{\otimes 2})\, O\, (V^\dagger)^{\otimes 2},
\label{eq:twirl_def}
\end{equation}
where $dV$ is the Haar measure on $U(d)$. By Schur--Weyl duality~\cite{collins2006integration,collins2003moments}, the commutant of $U(d)$ acting on two replicas is spanned by the identity and the swap, hence
\begin{equation}
\mathcal{M}^{(2)}_d[O] = a(O)\,\mathbb{I}_{d^2}+b(O)\,S_d ,
\label{eq:twirl_span}
\end{equation}
where $S_d$ swaps the two replicas of $\mathbb{C}^d$.

The coefficients follow from the two invariants preserved by the twirl:
\begin{equation}
\mathrm{Tr}\!\left[\mathcal{M}^{(2)}_d[O]\right]=\mathrm{Tr}[O],
\qquad
\mathrm{Tr}\!\left[S_d\,\mathcal{M}^{(2)}_d[O]\right]=\mathrm{Tr}[S_d\,O].
\label{eq:twirl_constraints}
\end{equation}
Solving for $a,b$ gives
\begin{equation}
a(O)=\frac{\mathrm{Tr}(O)-d^{-1}\mathrm{Tr}(S_d O)}{d^2-1},
\qquad
b(O)=\frac{\mathrm{Tr}(S_d O)-d^{-1}\mathrm{Tr}(O)}{d^2-1}.
\label{eq:ab_general}
\end{equation}

For a single qubit ($d=2$) we identify $S_d\equiv S$ and $\mathbb{I}_{d^2}\equiv \mathbb{I}_4$. In particular, for a pure state $\rho=\ket{\psi}\!\bra{\psi}$,
\begin{equation}
\mathcal{M}^{(2)}_2[\rho^{\otimes 2}]
=
\frac{1}{6}\,\mathbb{I}_4+\frac{1}{6}\,S,
\label{eq:pure_state_decomp}
\end{equation}
since $\mathrm{Tr}(\rho^{\otimes 2})=\mathrm{Tr}(S\rho^{\otimes 2})=\mathrm{Tr}(\rho^2)=1$.

\subsection*{Two-qubit Haar gate: local update matrix}
\label{app:two_qubit_update}

Consider a two-qubit unitary $U\in U(4)$ acting on $\mathbb{C}^4$.
In the folded two-replica picture, averaging the corresponding two-qubit gate over a unitary $2$-design
induces a two-site update on the basis
$\{\mathbb{I}_4\mathbb{I}_4,\mathbb{I}_4S,S\mathbb{I}_4,SS\}$.
By the general two-copy twirl formula [Eq.~\eqref{eq:ab_general} with $d=4$], the image of any two-site basis element
lies in $\mathrm{span}\{\mathbb{I}_4\mathbb{I}_4,SS\}$, and one obtains
\begin{equation}
\mathcal{M}^{(2)}_{4}[\mathbb{I}_4\mathbb{I}_4]=\mathbb{I}_4\mathbb{I}_4,\qquad
\mathcal{M}^{(2)}_{4}[SS]=SS,\qquad
\mathcal{M}^{(2)}_{4}[\mathbb{I}_4S]=\mathcal{M}^{(2)}_{4}[S\mathbb{I}_4]
=\frac{2}{5}\bigl(\mathbb{I}_4\mathbb{I}_4+SS\bigr).
\label{eq:twoqubit_rules}
\end{equation}

Therefore, the local update matrix in the non-normalized basis is
\begin{equation}
M_{\mathrm{Haar}}
=
\begin{pmatrix}
1 & \tfrac{2}{5} & \tfrac{2}{5} & 0 \\
0 & 0 & 0 & 0 \\
0 & 0 & 0 & 0 \\
0 & \tfrac{2}{5} & \tfrac{2}{5} & 1
\end{pmatrix}.
\label{eq:Mhaar}
\end{equation}

\subsection*{Noise and antinoise channels in the $\{\mathbb{I}_4,S\}$ basis}
\label{app:noise_updates}

We now derive the single-site update matrices for the noise channel and its inverse (antinoise) used in probabilistic error cancellation (PEC)~\cite{niroula2025error}; the corresponding comparison to TEM is discussed in the End Matter of the Letter.

Consider the single-qubit depolarizing channel
\begin{equation}
\mathcal{E}^{(1)}_{x,\tau}(\rho)
=
(1-q_{x,\tau})\,\rho
+
q_{x,\tau}\,\mathrm{Tr}(\rho)\,\frac{\mathbb{I}}{2}.
\label{eq:single_qubit_depol}
\end{equation}
On two replicas it acts as $\mathcal{E}_{x,\tau}\equiv \mathcal{E}^{(1)}_{x,\tau}\otimes \mathcal{E}^{(1)}_{x,\tau}$ on the operator space $(\mathbb{C}^2)^{\otimes 2}$.
Clearly,
\begin{equation}
\mathcal{E}_{x,\tau}(\mathbb{I}_4)=\mathbb{I}_4.
\end{equation}
To obtain $\mathcal{E}_{x,\tau}(S)$ it is convenient to use
\begin{equation}
S=\frac{1}{2}\left(\mathbb{I}\otimes \mathbb{I}+X\otimes X+Y\otimes Y+Z\otimes Z\right),
\label{eq:swap_pauli}
\end{equation}
so that the traceless Pauli--Pauli terms pick up a factor $(1-q_{x,\tau})^2$ under
$\mathcal{E}^{(1)}\otimes \mathcal{E}^{(1)}$, while the identity part remains invariant. This gives
\begin{equation}
\mathcal{E}_{x,\tau}(S)
=
(1-q_{x,\tau})^2\,S
+
\frac{1-(1-q_{x,\tau})^2}{2}\,\mathbb{I}_4.
\label{eq:E_on_S}
\end{equation}
Therefore, in the ordered basis $\{\mathbb{I}_4,S\}$ we obtain
\begin{equation}
M\!\left[\mathcal{E}_{x,\tau}(q_{x,\tau})\right]
=
\begin{pmatrix}
1 & \tfrac{1-(1-q_{x,\tau})^2}{2}\\
0 & (1-q_{x,\tau})^2
\end{pmatrix}.
\label{eq:ME_sm}
\end{equation}

The inverse map used in PEC is implemented by the antinoise channel
\begin{equation}
\mathcal{A}^{(1)}_{x,\tau}(\rho)
=
\eta\rho
+
(1-\eta)\,\mathrm{Tr}(\rho)\,\frac{\mathbb{I}}{2},
\qquad
\eta=\frac{1}{1-q_{a}}.
\label{eq:single_qubit_antinoise}
\end{equation}
Proceeding as above, one finds
\begin{equation}
\mathcal{A}_{x,\tau}(\mathbb{I}_4)=\mathbb{I}_4,
\qquad
\mathcal{A}_{x,\tau}(S)
=
(1-q_a)^{-2}\,S
+
\frac{1-(1-q_a)^{-2}}{2}\,\mathbb{I}_4,
\end{equation}
and hence
\begin{equation}
M\!\left[\mathcal{A}_{x,\tau}(q_a)\right]
=
\begin{pmatrix}
1 & \tfrac{1-(1-q_a)^{-2}}{2}\\
0 & (1-q_a)^{-2}
\end{pmatrix}.
\label{eq:MA_sm}
\end{equation}
This construction admits an effective RFIM-type parametrization. Accordingly, a two-site contribution to the statistical weight connecting neighboring sites $x$ and $y$ can be written as
$
\exp\!\Bigl[
-\Bigl(
J_\tau\, s_x s_y
+\bigl(r_{x,\tau}+u\bigr)\, s_x
+\bigl(r_{y,\tau}+u\bigr)\, s_y
\Bigr)
\Bigr],
$
where $
s_x = 1-2\alpha_x \in \{+1,-1\}
$, $J_\tau$ is set by $M_{\mathrm{Haar}}$, the random fields $r_{x,\tau}$ inherit their site dependence from $M[\mathcal{E}_{x,\tau}]$, and the uniform shift $u$ is inherited from $M[\mathcal{A}_{x,\tau}]$. $\alpha\in\{0,1\}^N$ labels the local basis element on each site, with $\alpha_x=0$ for $\mathbb{I}_4$ and $\alpha_x=1$ for $S$. In this parametrization, the zero-mean-field condition in Eq.~\eqref{eq:MF} is equivalently expressed as $\langle r_{x,\tau}+u\rangle=0$.

\subsection*{Diagrammatic mapping and tensor-network representation}
\label{app:diagrammatic_mapping}

The two-replica moment operators admit a convenient folded Choi representation, in which the averaged two-qubit gate is a four-leg tensor acting on the local $\{\mathbb{I}_4,S\}$ indices.
In this language one may write schematically
\begin{equation}
M_{\mathrm{Haar}}
=\mathbb{E}_U \left[ U \otimes U^* \otimes U \otimes U^* \right] .
\end{equation}
The corresponding local mapping used in Eq.~\eqref{eq:Mhaar} is summarized by the diagrammatic identity ~\cite{gao2024limitations,li2023entanglement}:
\begin{equation}
\label{eq:Haar_Map_Full}
\begin{tikzpicture}[baseline={(0, -0.5ex)}, x=0.62cm, y=0.62cm, font=\footnotesize]
    \node[font=\large] at (-2.2, 0) {$\mathbb{E}_U \Bigg[$};
    \node[font=\large] at ( 2.2, 0) {$\Bigg]$};

    \foreach \layer/\xoffset/\yoffset in {4/0.9/0.45, 3/0.6/0.15, 2/0.3/-0.15, 1/0/-0.45} {
        \pgfmathparse{int(mod(\layer,2))}
        \ifnum\pgfmathresult=1
            \def\fillcol{blockA} \def\isStar{0}
        \else
            \def\fillcol{blockB} \def\isStar{1}
        \fi

        \begin{scope}[shift={(\xoffset, \yoffset)}]
            \draw[black, line width=1.05pt] (-1.1, 0.25) -- (1.1, 0.25);
            \draw[black, line width=1.05pt] (-1.1,-0.25) -- (1.1,-0.25);
            \draw[fill=\fillcol, draw=black, line width=0.8pt, rounded corners=2pt] 
                (-0.65, -0.5) rectangle (0.65, 0.5);
            \ifnum\isStar=1
                \node[text=black, inner sep=0pt, font=\large] at (0.45, 0.35) {$*$};
            \fi
        \end{scope}
    }
\end{tikzpicture}
\quad = \quad
%
\begin{tikzpicture}[baseline={(0, -0.5ex)}, x=0.62cm, y=0.62cm]
    \tikzset{
      edge/.style={black, line width=1.05pt}
    }
    \def\rfilled{2.45pt}

    \coordinate (L) at (-0.5, 0); 
    \coordinate (R) at ( 0.5, 0);
    
    \coordinate (L_up)   at (-1.0,  0.866);
    \coordinate (L_down) at (-1.0, -0.866);
    \coordinate (R_up)   at ( 1.0,  0.866);
    \coordinate (R_down) at ( 1.0, -0.866);

    \draw[edge] (L) -- (R);        
    \draw[edge] (L) -- (L_up);     
    \draw[edge] (L) -- (L_down);   
    \draw[edge] (R) -- (R_up);     
    \draw[edge] (R) -- (R_down);   

    \fill[spinA] (L) circle (\rfilled);
    \fill[spinB] (R) circle (\rfilled);
\end{tikzpicture}
\quad = \quad
%
\begin{tikzpicture}[baseline={(0, -0.5ex)}, x=0.62cm, y=0.62cm, font=\footnotesize]
    \tikzset{
      haarBlock/.style={draw=black, fill=black!10, line width=1.05pt, rounded corners=2pt}
    }
    
    \draw[black, line width=1.05pt] (-1.5,  0.25) -- (0,  0.25);
    \draw[black, line width=1.05pt] (-1.5, -0.25) -- (0, -0.25);
    
    \draw[black, line width=1.05pt] (0,  0.25) -- (1.5,  0.25);
    \draw[black, line width=1.05pt] (0, -0.25) -- (1.5, -0.25);

    \node[haarBlock, minimum width=1.2cm, minimum height=0.9cm] at (0,0) {$M_{\mathrm{Haar}}$};
\end{tikzpicture}
\quad = \quad
%
\begin{tikzpicture}[baseline={(0, -0.5ex)}, x=0.62cm, y=0.62cm]
    \tikzset{
      tensO/.style={circle, draw=black, fill=black!10, line width=1.05pt, minimum size=3.0mm, inner sep=0pt},
      edge/.style={black, line width=1.05pt}
    }

    \coordinate (Top)    at (0,  0.433);
    \coordinate (Bottom) at (0, -0.433);

    \draw[edge] (Top) -- (Bottom);

    \draw[edge] (Top) -- (-0.75, 0.433); 
    \draw[edge] (Top) -- ( 0.75, 0.433); 
    \draw[edge] (Bottom) -- (-0.75, -0.433); 
    \draw[edge] (Bottom) -- ( 0.75, -0.433); 

    \node[tensO] at (Top) {};
    \node[tensO] at (Bottom) {};
\end{tikzpicture}
\end{equation}

 where the blue and orange blocks denote $U$ and $U^*$ in the folded two-replica circuit. In the associated vertex model, each spacetime site carries a replica-permutation variable $g\in S_2=\{e,(12)\}$ ~\cite{weingarten1978asymptotic,collins2022weingarten}, equivalently labeled by the basis elements $\mathbb{I}_4$ and $S$. Haar-averaged two-qubit gates generate the nontrivial two-site interaction encoded in $M_{\mathrm{Haar}}$ [Eq.~\eqref{eq:Mhaar}], while the remaining legs implement the wiring (index contraction) between neighboring circuit elements. Viewed as a tensor, $M_{\mathrm{Haar}}$ is a two-site MPO building block; assembling these blocks according to the gate schedule yields the MPO for a full one-step transfer operator acting on an MPS over the spatial sites.

For a product initial state $\rho_{\mathrm{ini}}=(\ket{\psi_i}\!\bra{\psi_i})^{\otimes N}$, each site contributes the local two-replica operator \eqref{eq:pure_state_decomp},
\begin{equation}
\bar{\rho}^{(2)}_{\mathrm{ini}}
=
\left(\frac{1}{6}\,\mathbb{I}_4+\frac{1}{6}\,S\right)^{\otimes N}.
\label{eq:rho_ini_app}
\end{equation}
Similarly, the final boundary state is a product functional determined by the observable insertion.

\begin{figure}[t]
  \centering
  \includegraphics[width=\linewidth]{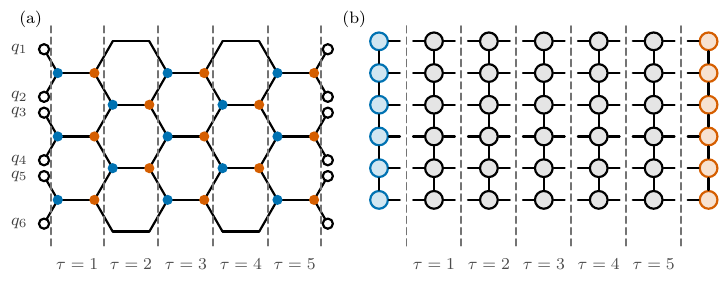}
\caption{
(a) Diagrammatic mapping of an ideal 1D brickwork Haar-random circuit with open spatial boundary conditions onto a $(1\!+\!1)$D statistical model on the honeycomb spacetime lattice. The labels $q_1,\ldots,q_6$ indicate the site ordering, and the opposite boundary denotes the final contraction. Local identities are given in Eqs.~\eqref{eq:Haar_Map_Full} and \eqref{eq:Boundary_Mappings_OneLine}.
(b) Tensor-network representation of the associated transfer process. The coefficient vector $|P_{\tau}\rangle$ at $\tau=0$ is represented by a blue MPS. Each circuit layer applies a gray MPO representing the transfer matrix $M[\mathcal{U}_\tau]$. Second-moment observables are evaluated by contracting the network with the appropriate orange boundary MPS at the final time.}
  \label{stats_mapp_TNSl}
\end{figure}

The boundary mapping used in Figure~\ref{stats_mapp_TNSl} is:
\begin{equation}
\label{eq:Boundary_Mappings_OneLine}
\begin{tikzpicture}[baseline={(0, -0.11)}, x=0.62cm, y=0.62cm, font=\footnotesize]
    \node[font=\large, anchor=east] at (-3.0, 0) {$\mathbb{E}_U \Bigg[$};
    \node[font=\large, anchor=west] at ( 1.8, 0) {$\Bigg]$};
    \foreach \layer/\xoffset/\yoffset in {4/0.9/0.45, 3/0.6/0.15, 2/0.3/-0.15, 1/0/-0.45} {
        \pgfmathparse{int(mod(\layer,2))}
        \ifnum\pgfmathresult=1 \def\fillcol{blockU} \def\isStar{0}
        \else \def\fillcol{blockUstar} \def\isStar{1} \fi
        \begin{scope}[shift={(\xoffset, \yoffset)}]
            \draw[black, line width=1.05pt] (-2.3, 0) -- (-1.3, 0); 
            \draw[black, line width=1.05pt] (-0.3, 0) -- (0.7, 0);
            \draw[fill=black!10, draw=black, line width=0.8pt, rounded corners=2pt] 
                (-2.9, -0.4) rectangle (-1.7, 0.4);
            \node at (-2.3, 0) {$\ket{\psi_i}$};
            \draw[fill=\fillcol, draw=black, line width=0.8pt, rounded corners=2pt] 
                (-1.3, -0.4) rectangle (-0.3, 0.4);
            \ifnum\isStar=1 \node[text=black, inner sep=0pt, font=\small] at (-0.5, 0.2) {$*$}; \fi
        \end{scope}
    }
\end{tikzpicture}%
\; = \;%
\begin{tikzpicture}[baseline={(0, -0.11)}, x=0.62cm, y=0.62cm]
    \def\rfilled{2.45pt}
    \tikzset{edge/.style={black, line width=1.05pt}}
    \draw[edge] (0,0) -- (1.0, 0);
    \draw[black, line width=1.05pt, fill=white] (0,0) circle (\rfilled);
    \node[anchor=east, text=black!75, font=\footnotesize] at (-0.15, 0) {$i$};
\end{tikzpicture}%
\; = \;%
\begin{tikzpicture}[baseline={(0, -0.11)}, x=0.62cm, y=0.62cm]
    \tikzset{tensL/.style={circle, draw=spinA, fill=spinA!18, line width=1.05pt, minimum size=3.0mm, inner sep=0pt},
             edge/.style={black, line width=1.05pt}}
    \coordinate (Center) at (0,0);
    \draw[edge] (Center) -- (0,  0.8);
    \draw[edge] (Center) -- (0, -0.8);
    \draw[edge] (Center) -- (1.0, 0);
    \node[tensL] at (Center) {};
\end{tikzpicture}%
%
,\qquad%
%
\begin{tikzpicture}[baseline={(0, -0.11)}, x=0.62cm, y=0.62cm, font=\footnotesize]
    \node[font=\large, anchor=east] at (-2.45, 0) {$\mathbb{E}_U \Bigg[$};
    \node[font=\large, anchor=west] at ( 2.3, 0) {$\Bigg]$};
    \foreach \layer/\xoffset/\yoffset in {4/0.9/0.45, 3/0.6/0.15, 2/0.3/-0.15, 1/0/-0.45} {
        \pgfmathparse{int(mod(\layer,2))}
        \ifnum\pgfmathresult=1 \def\fillcol{blockU} \def\isStar{0}
        \else \def\fillcol{blockUstar} \def\isStar{1} \fi
        \begin{scope}[shift={(\xoffset, \yoffset)}]
            \draw[black, line width=1.05pt] (-2.3, 0) -- (-1.3, 0); 
            \draw[black, line width=1.05pt] (-0.3, 0) -- (0.1, 0);
            \draw[fill=\fillcol, draw=black, line width=0.8pt, rounded corners=2pt] 
                (-1.3, -0.4) rectangle (-0.3, 0.4);
            \ifnum\isStar=1 \node[text=black, inner sep=0pt, font=\small] at (-0.5, 0.2) {$*$}; \fi
            \draw[fill=black!10, draw=black, line width=0.8pt, rounded corners=2pt] 
                (0.1, -0.4) rectangle (1.3, 0.4);
            \node at (0.7, 0) {$\bra{\psi_f}$};
        \end{scope}
    }
\end{tikzpicture}%
\; = \;%
\begin{tikzpicture}[baseline={(0, -0.11)}, x=0.62cm, y=0.62cm]
    \def\rfilled{2.45pt}
    \tikzset{edge/.style={black, line width=1.05pt}}
    \draw[edge] (0,0) -- (-1.0, 0);
    \draw[black, line width=1.05pt, fill=white] (0,0) circle (\rfilled);
    \node[anchor=west, text=black!75, font=\footnotesize] at (0.2, 0) {$f$};
\end{tikzpicture}%
\; = \;%
\begin{tikzpicture}[baseline={(0, -0.11)}, x=0.62cm, y=0.62cm]
    \tikzset{tensR/.style={circle, draw=spinB, fill=spinB!18, line width=1.05pt, minimum size=3.0mm, inner sep=0pt},
             edge/.style={black, line width=1.05pt}}
    \coordinate (Center) at (0,0);
    \draw[edge] (Center) -- (0,  0.8);
    \draw[edge] (Center) -- (0, -0.8);
    \draw[edge] (Center) -- (-1.0, 0);
    \node[tensR] at (Center) {};
\end{tikzpicture}
\end{equation}

Second-moment observables are obtained by contracting the final two-replica state with product boundary functionals. In the $\{\mathbb{I}_4,S\}$ basis these are represented by
\begin{equation}
\boldsymbol{v}_{\mathrm{Tr}}=\begin{pmatrix}4\\2\end{pmatrix},
\qquad
\boldsymbol{v}_{\mathrm{p}}=\begin{pmatrix}2\\4\end{pmatrix},
\label{eq:boundary_vectors_app}
\end{equation}
corresponding to taking a trace or inserting a swap, respectively. For a subsystem $a$, the circuit-averaged purity is obtained by contracting with
$
\bigotimes_{x\in a}\boldsymbol{v}_{\mathrm{p}}
\;\otimes\;
\bigotimes_{x\notin a}\boldsymbol{v}_{\mathrm{Tr}}
$.

With the local tensors specified above, the two-replica evolution over one elementary transfer step $\tau$ is represented as an MPO acting on the coefficient vector in the basis $\{\mathbb{I}_4,S\}^{\otimes N}$. We denote the gate-only transfer MPO by $M[\mathcal{U}_\tau]$, assembled from the two-qubit tensors $M_{\mathrm{Haar}}$ placed on the gate schedule. Starting from the product initial state in Eq.~\eqref{eq:rho_ini_app}, we represent the coefficient vector of $\bar{\rho}^{(2)}_\tau$ as an MPS $|P_\tau\rangle$ with components $P_\tau(\alpha)$. A one-step update is implemented by $|P_{\tau+1}\rangle=M[\mathcal{U}_\tau]|P_\tau\rangle$, and second-moment observables are obtained by contracting $|P_\tau\rangle$ with the corresponding product boundary functional at the final time $\tau$. Figure~\ref{stats_mapp_TNSl} depicts this ideal one-dimensional mapping and its tensor-network representation.

For error-mitigated evolution, the step update is dressed by inserting the on-site channel tensors,
$\mathbb{M}_\tau \equiv M[\mathcal{A}_\tau]\;M[\mathcal{E}_\tau]\;M[\mathcal{U}_\tau]$,
with the same channel-composition convention in Eq.~\eqref{eq:transfer_update}. Numerically, $\mathbb{M}_\tau$ is applied using a TEBD decomposition into two-site updates, followed by truncation to the chosen bond dimension.

To quantify the entanglement structure of the coefficient MPS, we introduce the $\ell_2$-normalized coefficient vector
$|\tilde P_\tau\rangle\equiv |P_\tau\rangle/\|P_\tau\|_2$. For a
subsystem $R$ in the chosen MPS ordering, with complement $\bar R$, we define
\begin{equation}
\begin{split}
\tilde{\varrho}^{(2)}_R(\tau)
&\equiv
\operatorname{Tr}_{\bar R}\!\left[
|\tilde P_\tau\rangle\langle\tilde P_\tau|
\right],\\
S_{\mathrm{MPS}}(\tau)
&=
-\operatorname{Tr}\!\left[
\tilde{\varrho}^{(2)}_R(\tau)
\log\tilde{\varrho}^{(2)}_R(\tau)
\right].
\end{split}
\label{eq:Smps_supp}
\end{equation}
This is the bipartite MPS entanglement entropy shown in Fig.~\ref{fig:1d_S_MPS_vertical}(a) of the Letter.

For two-dimensional geometries we employ a snake (zig-zag) ordering consistent with the cylindrical boundary conditions shown in Fig.~\ref{fig:supp_gate_schedules}. For the all-to-all architecture, each layer is a disjoint random matching; we apply the same fixed-$\chi$ TEBD update to the corresponding set of long-range two-site tensors.

\begin{figure}
  \centering
  \includegraphics[width=0.5\linewidth]{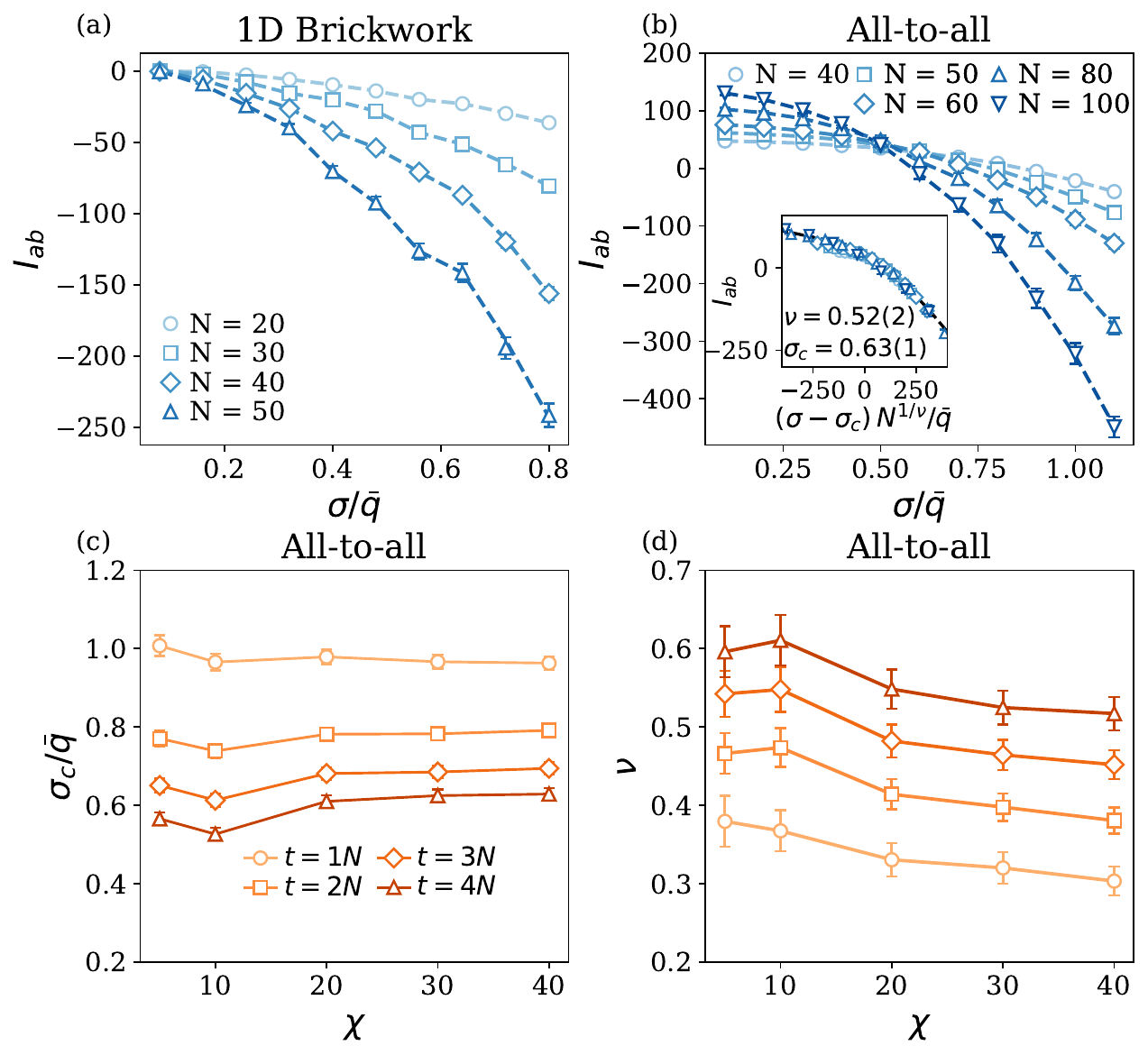}
\caption{Mutual-information diagnostic for the 1D and all-to-all
benchmarks.
(a),(b) Disorder dependence of the circuit-averaged mutual information
$I_{ab}$ at layer $t=4N$ and fixed bond dimension $\chi=40$ for
(a) quenched 1D and (b) annealed all-to-all circuits. Colored lines
are guides to the eye.
The inset in (b) displays the critical data collapse near the
mitigation threshold; the black dashed curve denotes the fitted
scaling function.
(c),(d) Extracted threshold location $\sigma_c/\bar q$ and critical
exponent $\nu$, respectively, as functions of bond dimension $\chi$
for the all-to-all geometry.
Error bars in (c) and (d) denote finite-size-scaling fit uncertainties,
and the quoted uncertainties in the inset of (b) are obtained from
the covariance matrix of the nonlinear collapse fit.}
  \label{fig:supp_1d_all_I}
\end{figure}

\begin{figure}
  \centering
  \includegraphics[width=0.5\linewidth]{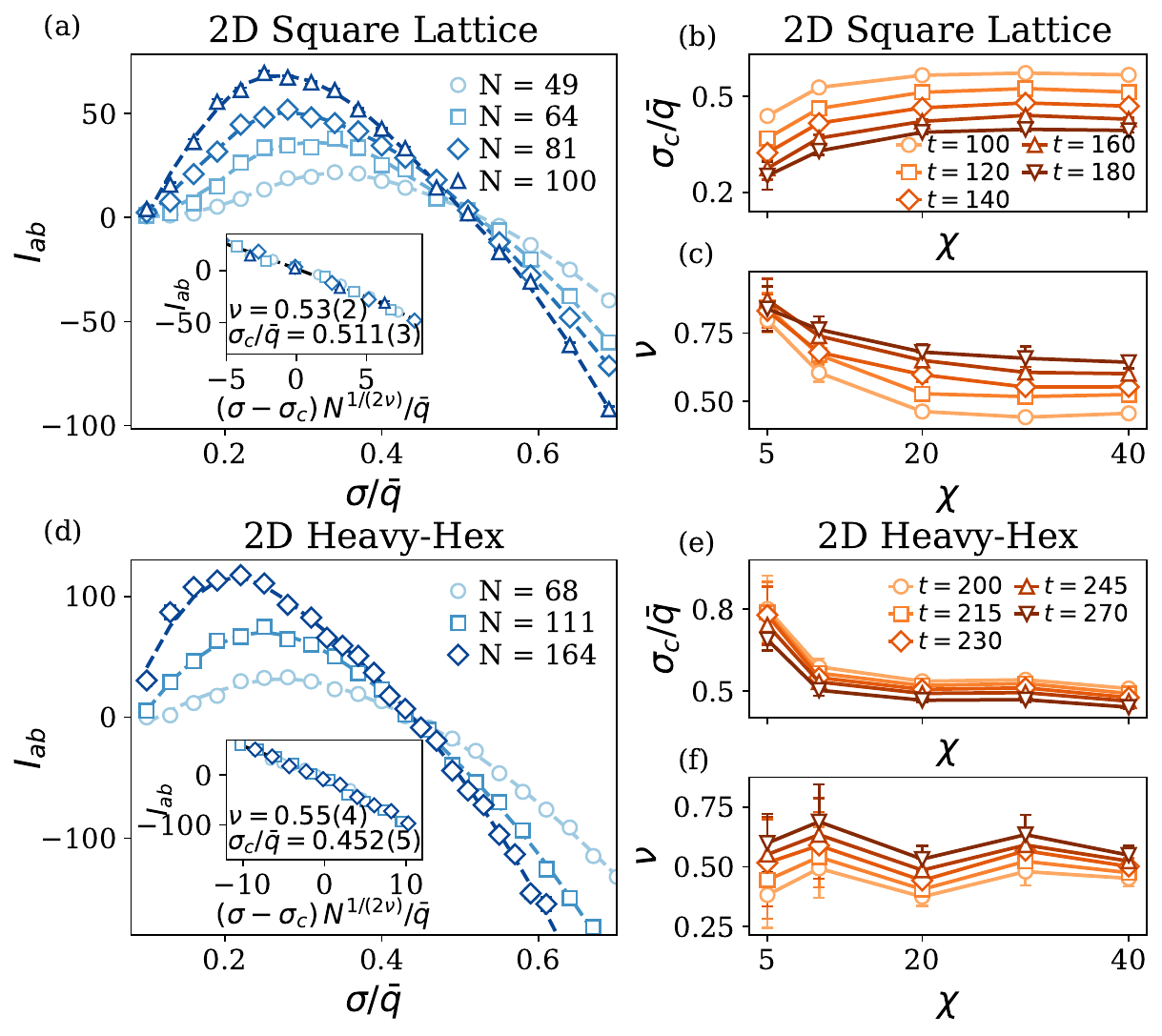}
\caption{Mutual-information diagnostic for 2D architectures.
(a) Disorder dependence of $I_{ab}$ for the square lattice at
$t=120$ and $\chi=40$.
(d) The corresponding heavy-hex results at $t=270$ and $\chi=40$.
Colored lines are guides to the eye.
Insets in (a) and (d) show the finite-size-scaling collapse versus
$(\sigma-\sigma_c)N^{1/(2\nu)}/\bar q$; black dashed curves denote
the fitted scaling functions.
(b),(c) Bond-dimension dependence of the extracted
$\sigma_c/\bar q$ and $\nu$ for the square lattice.
(e),(f) The corresponding bond-dimension dependence for the
heavy-hex lattice.
Error bars denote finite-size-scaling fit uncertainties, while
quoted inset uncertainties are obtained from the covariance matrix
of the nonlinear collapse fit.}
 \label{fig:supp_2d_I}
\end{figure}

\begin{figure*}
  \centering
  \includegraphics[width=\linewidth]{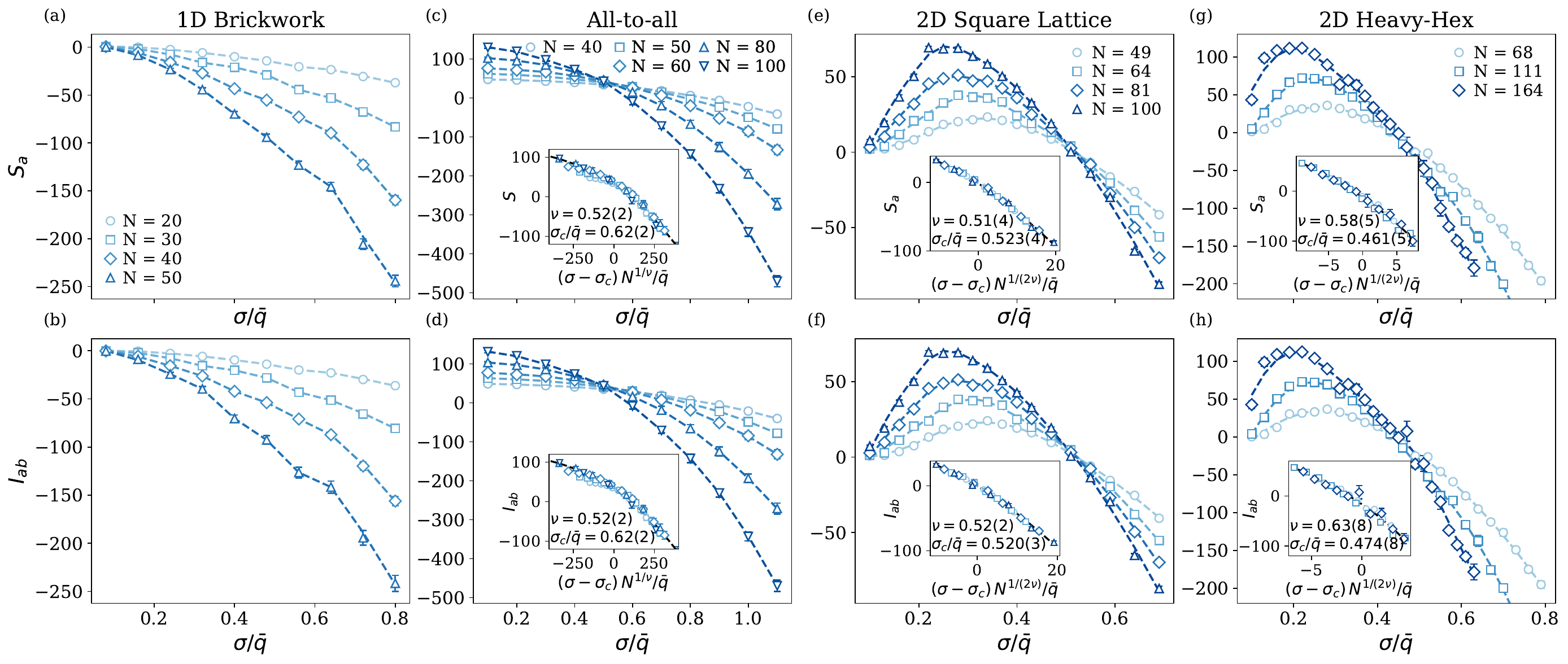}
  \caption{Intermediate-bond-dimension check at $\chi=30$.
  Single-site R\'enyi-2 entropy $S_a$ (top row) and two-point mutual
  information $I_{ab}$ (bottom row) versus disorder strength
  $\sigma/\bar q$ for (a),(b) the 1D brickwork geometry with quenched
  disorder at $t=4N$; (c),(d) the all-to-all geometry with annealed
  disorder at $t=4N$; (e),(f) the square lattice with annealed disorder
  at $t=120$; and (g),(h) the heavy-hex lattice with annealed disorder
  at $t=270$. Colored dashed lines are guides to the eye. Insets in
  (c)--(h) show the finite-size-scaling collapses; black dashed curves
  denote the fitted scaling functions.}
  \label{fig:supp_convergence_chi30}
\end{figure*}

\subsection*{Finite-size scaling procedure}

We extract the threshold location and critical exponent from finite-size-scaling collapses of the disorder-averaged observables $O\in\{S_a,I_{ab}\}$ in the vicinity of the crossing region. For each geometry, depth, and bond dimension $\chi$, we first estimate
the crossing from the intersection of the interpolated finite-size
curves. We then perform the collapse using the sampled
disorder values closest to this estimate, using the same set of system sizes shown in the corresponding panel. For the all-to-all geometry, the scaling ansatz is
\begin{equation}
O(N,\sigma)=F(X),
\qquad
X=\left(\frac{\sigma}{\bar q}-\frac{\sigma_c}{\bar q}\right)N^{1/\nu},
\label{eq:fss_ansatz_sm}
\end{equation}
where $\sigma_c/\bar q$ is the critical disorder and $\nu$ is the critical exponent. For the two-dimensional square and heavy-hex geometries, we instead use $N^{1/(2\nu)}$ in the definition of $X$.

We approximate the scaling function by a cubic polynomial,
\begin{equation}
F(X)\approx c_0+c_1X+c_2X^2+c_3X^3,
\label{eq:fss_poly_sm}
\end{equation}
so that the fit parameters are $(\sigma_c/\bar q,\nu,c_0,c_1,c_2,c_3)$.

Because the computational cost increases with bond dimension, each
disorder-averaged data point is obtained from $10^4$ independent
circuit realizations for $20\leq\chi\leq40$ and from $10^5$
realizations for $\chi<20$. These sample counts are used for all geometries. For the
all-to-all geometry, each circuit realization additionally includes an
independently sampled sequence of random disjoint matchings, with one
matching drawn at each complete circuit layer.

The collapse fits are performed by weighted nonlinear least squares with weights $w_i=1/\delta O_i^2$, where $\delta O_i$ denotes the standard error of the disorder-averaged data point at fixed $(N,\sigma)$. The quoted uncertainties in $\sigma_c/\bar q$ and $\nu$ are obtained from the covariance matrix of the nonlinear fit. The same procedure is used for both $S_a$ and $I_{ab}$.

\subsection*{Mutual information diagnostics}
\label{app:mi_truncation}

In addition to the single-qubit R\'enyi-2 entropy $S_a$, we use the two-point R\'enyi-2 mutual information $I_{ab}\equiv S_a+S_b-S_{ab}$ as an independent diagnostic of the error-mitigation transition. For the 1D and all-to-all geometries we choose single-qubit subsystems $a$ and $b$ to be separated by $N/2$ sites along the MPS ordering. For the 2D square and heavy-hex geometries, we choose $a$ at a corner of the finite lattice and $b$ at the site closest to the geometric center. We have checked several alternative choices of the probe sites with different $a$-$b$ separations and find qualitatively similar disorder dependence and crossing behavior.

As an independent check of  the numerical benchmarks, we evaluate $I_{ab}$ at depth $t=4N$. For the quenched-disorder 1D brickwork geometry, $I_{ab}$ does not exhibit a stable crossing for any $\sigma/\bar q>0$ (Fig.~\ref{fig:supp_1d_all_I}(a)), consistent with the absence of a finite mitigation threshold inferred from $S_a$. For the annealed-disorder all-to-all architecture, $I_{ab}$ exhibits a clear crossing that sharpens with increasing $N$ (Fig.~\ref{fig:supp_1d_all_I}(b)). A finite-size scaling collapse yields $\sigma_c/\bar q = 0.63(1)$ and $\nu = 0.52(2)$, consistent with the $S_a$ analysis within uncertainties (Fig.~\ref{fig:crit_vs_chi}(c)--(d)).

For the 2D square and heavy-hex architectures, we evaluate $I_{ab}$ using the same schedules and depths as the primary $S_a$ analysis (Fig.~\ref{fig:supp_2d_I}(a) and (d)). Independent finite-size scaling collapses yield $\sigma_c/\bar q = 0.511(3)$ and $\nu =  0.53(2)$ for the square lattice, and $\sigma_c/\bar q = 0.452(5)$ and $\nu = 0.55(4)$ for the heavy-hex lattice, consistent with the $S_a$ estimates within fit uncertainties. Both extracted quantities are stable at the largest bond dimensions,
although $\nu$ shows greater finite-$\chi$ variation at smaller $\chi$ [Fig.~\ref{fig:supp_2d_I}(c)--(f)], mirroring the behavior observed for $S_a$. This difference is expected because $\sigma_c/\bar q$ is determined primarily by the crossing location, whereas $\nu$ is inferred from the size-dependent slopes and is therefore more sensitive to finite-$\chi$ distortions of the scaling collapse.

As an additional consistency check, Fig.~\ref{fig:supp_convergence_chi30} shows results at an intermediate bond dimension $\chi=30$ across all geometries. The all-to-all and 2D results retain clearly visible crossings in both $S_a$ and $I_{ab}$, while the 1D results continue to show no stable crossing at nonzero disorder. Finite-$\chi$ truncation introduces a systematic drift in the crossing
location and, at smaller bond dimensions, enhances the
realization-to-realization fluctuations of the observables, leading to
larger disorder-sampling uncertainties and less stable fits.

\subsection*{Depth dependence of the finite-depth thresholds}
\label{app:depth_dependence}

\begin{figure*}[t]
  \centering
  \includegraphics[width=0.92\linewidth]
  {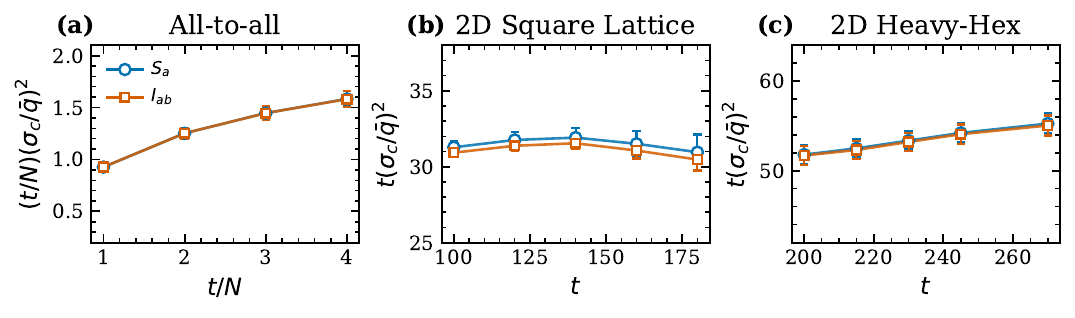}
  \caption{Scaled finite-depth thresholds at $\chi=40$.
  (a) $(t/N)(\sigma_c/\bar q)^2$ versus $t/N$ for the all-to-all
  geometry.
  (b),(c) $t(\sigma_c/\bar q)^2$ versus $t$ for the square and
  heavy-hex geometries, respectively.
  Blue circles and orange squares denote thresholds extracted from
  $S_a$ and $I_{ab}$.
  Error bars are propagated from covariance estimates of $\sigma_c/\bar q$ obtained in the corresponding finite-size-scaling
fits.}
  \label{fig:supp_depth_scaling}
\end{figure*}

We examine the depth dependence of the thresholds extracted at
$\chi=40$, motivated by the scaled-depth phase diagram in
Fig.~\ref{fig:circuits}(a). The thresholds obtained independently
from $S_a$ and $I_{ab}$ remain consistent throughout the simulated
depth range [Fig.~\ref{fig:supp_depth_scaling}]. For the square
lattice, $t(\sigma_c/\bar q)^2$ is approximately depth independent
within the fit uncertainties, whereas the heavy-hex results exhibit
a modest residual drift. The all-to-all quantity
$(t/N)(\sigma_c/\bar q)^2$ displays stronger finite-depth evolution
over $t/N=1,\ldots,4$. These results characterize the
architecture-dependent finite-depth phase boundaries reported in
the Letter.

\subsection*{Additional convergence data for truncation paths}
\label{app}
\begin{figure}
  \centering
  \includegraphics[width=\linewidth]{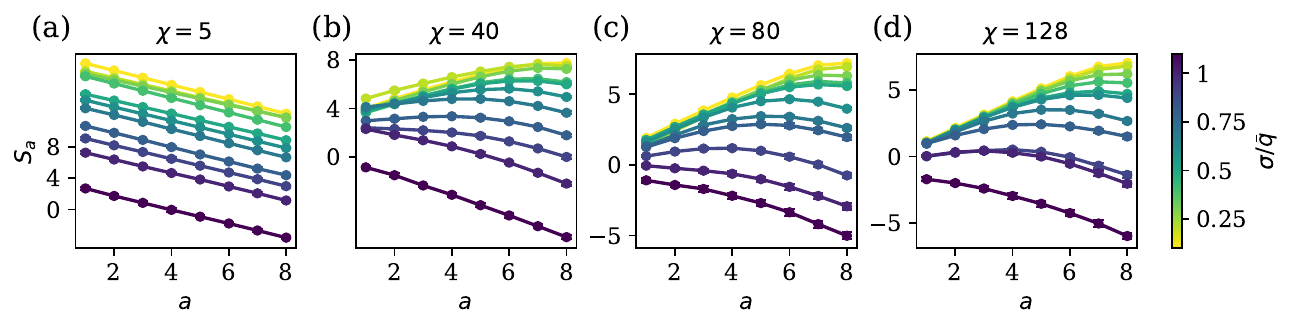}
  \caption{Convergence data for the renormalization-free protocol in the non-normalized $\{\mathbb{I}_4,S\}^{\otimes N}$ basis. Scaling of the R\'enyi-2 entropy $S_a$ with subsystem size for an $N=16$ all-to-all circuit with annealed disorder at depth $t=3N$. Increasing $\chi$ yields systematic convergence toward exact simulation for this small system while preserving visible disorder dependence already at relatively small bond dimension.}
  \label{fig:supp_truncation_main}
\end{figure}

\begin{figure}
  \centering
  \includegraphics[width=\linewidth]{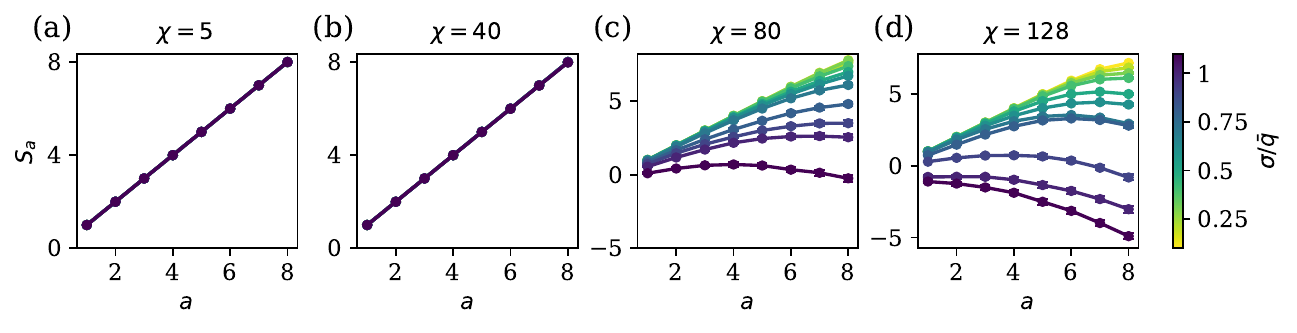}
  \caption{Convergence data for the normalized-basis protocol with post-truncation renormalization. At small $\chi$, the additional basis/normalization bias strongly reduces disorder sensitivity and delays convergence relative to Fig.~\ref{fig:supp_truncation_main}.}
  \label{fig:supp_truncation_alt}
\end{figure}

\begin{figure}[t]
  \centering
  \includegraphics[width=0.5\linewidth]{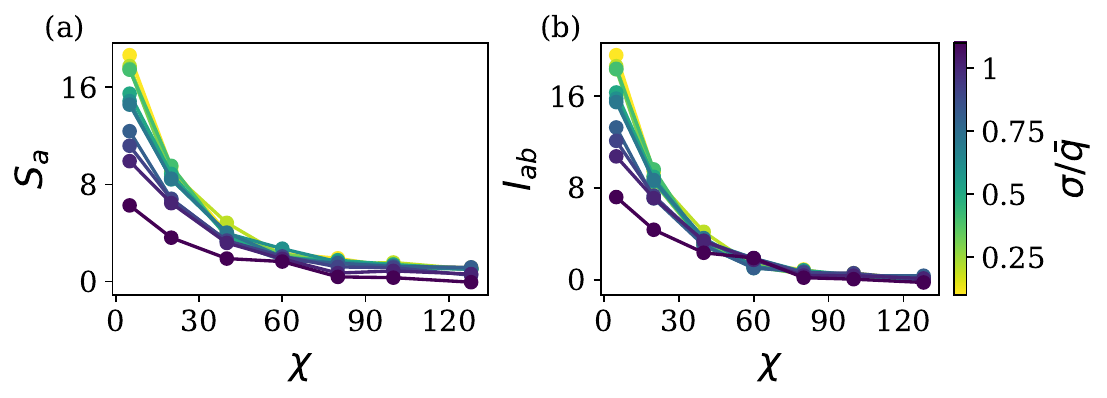}\par
  \vspace{-1.2mm}
  \includegraphics[width=0.5\linewidth]{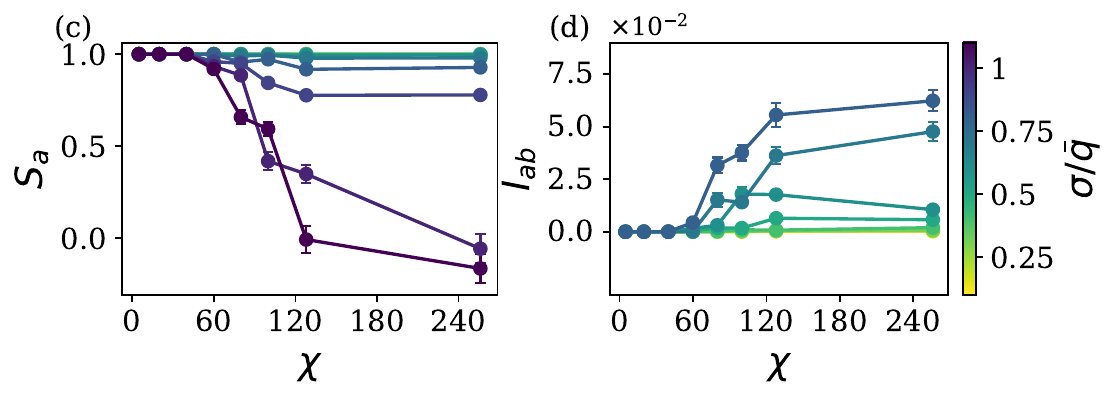}
\caption{Bond-dimension convergence of the single-qubit R\'enyi-2 entropy $S_a$ and two-point mutual information $I_{ab}$ for an $N=16$ all-to-all circuit with annealed disorder at depth $t=3N$. Panels (a)--(b) use the renormalization-free protocol in the non-normalized basis; panels (c)--(d) use the normalized-basis protocol with post-truncation renormalization.}
  \label{fig:SI_vs_chi_two_rows}
\end{figure}

End Matter summarizes the conceptual role of the truncation path in the fixed-$\chi$ protocol. Here we provide the corresponding bond dimension dependence data. Figures~\ref{fig:supp_truncation_main} and~\ref{fig:supp_truncation_alt} compare the subsystem-size scaling of $S_a$ for the renormalization-free protocol in the non-normalized basis and for the normalized-basis protocol with post-truncation renormalization. Fig.~\ref{fig:SI_vs_chi_two_rows} shows the corresponding bond-dimension dependence of $S_a$ and $I_{ab}$.

\end{document}